\documentclass{jfm}
\usepackage{graphicx}
\usepackage{epstopdf, epsfig}
 \usepackage{amssymb}
\usepackage{comment}
\usepackage{amsmath}
\usepackage{subcaption}
\usepackage{placeins}
\usepackage{xcolor}
 \usepackage{hyperref}

\newcommand{\D}{\partial}

\newcommand{\bv}[1]{\boldsymbol{#1}}

\newcommand{\uvec}{\bv{u}}

\newcommand{\dnsfig}[1]{\IfFileExists{#1}{\includegraphics[width=\linewidth]{#1}}{\fbox{\parbox[c][5cm][c]{0.95\linewidth}{\centering\color{red} Insert figure: \detokenize{#1}}}}}

\shorttitle{roll--streak dynamo instabilities and equilibria in MHD Couette flow}
\shortauthor{Eojin Kim and Brian F. Farrell}

\title{Statistical state dynamics of composite velocity/magnetic-field roll-streak instabilities and their finite-amplitude equilibria in turbulent MHD Couette flow}

\author{Eojin Kim\aff{1}
  \corresp{\email{ekim@g.harvard.edu}} \and
  Brian F. Farrell\aff{2}
	}
\affiliation{\aff{1}Department of Astrophysical Sciences, Princeton University, Princeton, NJ 08544, USA
\aff{2}Department of Earth and Planetary Sciences, Harvard University, Cambridge, MA 02139, USA}

\begin{document}

\maketitle

\begin{abstract}

The roll streak structure (RSS) is ubiquitous in shear flow turbulence and is fundamental to the dynamics of the self-sustaining process (SSP) maintaining the turbulent state.  Formation of the RSS in shear flow suggests an underlying instability that has recently been  identified using statistical state dynamics (SSD).  The mathematical analogy between the Navier--Stokes equation and the magnetic induction equation suggests investigating the possibility of a corresponding SSD instability in magnetohydrodynamic (MHD) shear flow.  In this work we both verify that such a composite velocity-magnetic field instability arises in the MHD SSD of Couette flow and also connect this instability to its nonlinear equilibria. In the SSD these equilibria correspond to either a steady fixed point or to chaotic, self-sustaining turbulent states. These instabilities and equilibria provide an analytical characterization for the formation and equilibration of large scale coherent dynamos in shear flow.

\end{abstract}



\section{Introduction}\label{sec:intro}
The roll streak structure (RSS) is commonly observed in nonconducting wall-bounded shear flow turbulence \citep{Kline 1962, Smith 1983}.
The formation of the RSS in these flows was originally attributed to the lift-up mechanism \citep{Ellingsen 1975, Gustavsson 1991}.   Because the RSS is not unstable in the  linearized  Navier-Stokes equation (NSE) for shear flow it could not be explained as an unstable mode.  However,  an analytical explanation for its ubiquity in wall bounded shear flow was  obtained using optimal perturbation analysis by showing that the RSS is the structure of optimal transient growth \citep{Butler-Farrell-1992, Reddy 1993}.  Subsequently, it was found that the RSS is unstable in turbulent shear flow when the NSE is written in statistical state dynamics (SSD) form \citep{Farrell-Ioannou-2012,Farrell-Ioannou-2016-bifur}. In SSD the dynamics is written for the statistics of the turbulence, its cumulants, rather than for a single realization.  In the SSD of shear flows, an unstable RSS arises by a mechanism in which Reynolds stresses associated with a weakly maintained unstructured background field of turbulence are systematically organized by the mean flow (first cumulant) perturbation component of the RSS mode to produce perturbation correlations (solved for by the second cumulant) properly oriented to force roll structures that in turn drive the lift-up mechanism underlying the streak component of the modal instability. These SSD modal instabilities with RSS form arise by the perturbation analog of the familiar self-sustaining process (SSP) that maintains turbulence in shear flow \citep{Hamilton-Kim-Waleffe-1995, Waleffe 1997}. At finite amplitude these modes continue to and equilibrate as either fixed-point RSS structures \citep{Farrell-Ioannou-2016-bifur} or may, as a function of parameters, transition to time-dependent self-sustaining turbulent states \citep{Farrell-Ioannou-2012}. These SSD RSS provide an analytical characterization of both the perturbation instability in weakly turbulent shear flows and of the finite amplitude statistical mean turbulent states of shear flow.

Because the governing MHD equations couple the momentum and magnetic induction equations, it follows that the coupled system could support nonlinear SSD instabilities, finite-amplitude statistical equilibria, and roll--streak (RSS) turbulence analogous to those observed in wall-bounded shear flows. Here, we demonstrate the existence of these coupled MHD states. A distinction in the dynamo mechanism identified in the SSD formulation of MHD in shear flow compared to dynamo mechanisms found in MHD formulations using traditional state variables is that the organization of Reynolds and Maxwell stresses in the SSD produces the dynamo-inducing feedback by which the composite MHD RSS is destabilized by background turbulence. Our analytical framework builds on generalized stability theory (GST) \citep{Farrell 1996a, Farrell 1996b} and the extension of GST to coherent structure formation in both nonconducting shear flows \citep{Butler-Farrell-1992, Farrell-Ioannou-1993-stoch} and conducting shear flows \citep{Farrell 1999a, Farrell 1999b}. GST was subsequently extended to study the dynamics of statistical states, i.e.\ SSD, in nonconducting flows \citep{Farrell-Ioannou-2003, Srinivasan-Young-2012, Parker-Krommes-2013, Constantinou-2014, Constantinou-2016} and conducting flows \citep{Farrell-Ioannou-2009-plasmas, Constantinou 2018, Parker-Constantinou-2019}.

According to Cowling's theorem \citep{Cowling 1933}, a two-dimensional mean magnetic field cannot sustain dynamo action through mean shear alone unless a net poloidal field is externally imposed (e.g., a uniform cross-stream field $\overline{B}_y=\mathrm{constant}$). Without an imposed field, the mean shear tilting of the poloidal field generates a toroidal field but the poloidal field decays unless it is actively regenerated. Classical mean-field models overcome this decay by invoking an $\alpha$-effect in which a kinematic parameterization rotates a part of the toroidal field into the poloidal direction \citep{Steenbeck 1966, Moffatt 1978, Krause 1980, Brandenburg 2005}. In this work, we identify a fundamentally distinct mechanism: a fluctuation feedback in which the SSD organizes velocity and magnetic field fluctuations to produce three distinct mean-field feedbacks: Reynolds stresses, Maxwell stresses, and a turbulent electromotive force (EMF). While the Reynolds and Maxwell stresses sustain the velocity field component of the roll--streak structure, the turbulent EMF plays an essential role in sustaining the coherent magnetic field component. While this feedback is analogous to the $\alpha$-effect in circumventing Cowling's theorem, it is mechanistically distinct and entails feedback by an explicitly resolved turbulent EMF. Importantly, once transition to turbulence occurs, the small scale turbulence parameterization required for RSS instability in the SSD can be removed, after which the turbulence self-sustains. Self-sustaining turbulence and large-scale self-sustaining dynamos have been found previously to occur in DNS of both Couette and Keplerian shear flow. In those studies a relation to subcritical transition, and an analogy with the self-sustaining process (SSP) in hydrodynamic flows were invoked \citep{Rincon 2007, Riols 2013, Nauman 2017}. In this work, we have verified the SSP in the SSD of MHD Couette turbulence, verified that the classical Couette velocity SSP exists in MHD Couette turbulence and that, as a function of Prandtl number, a self-sustaining dynamo also exists. The analytically characterized SSD RSS instability is shown in this work to lead to supercritical transition to a Couette MHD turbulent state that is self-sustaining and maintains a large scale dynamo. In summary, analysis using the SSD formulation of MHD turbulence in Couette flow reveals that a composite velocity--magnetic field RSS arises from a SSD instability. Equilibration of this instability leads either to a steady fixed point statistical state or to a chaotic statistical state.   Both of these statistical states can correspond to large-scale coherent dynamos and the latter  can be self-sustaining.

\section{Governing equations and boundary conditions}\label{sec:pimhd}

We study the dynamics of a turbulent, incompressible, conducting Couette flow with coordinates $x$, $y$, $z$ in the streamwise, cross-stream, and spanwise directions, respectively, and corresponding unit vectors $\hat{i}$, $\hat{j}$ and $\hat{k}$. The governing equations are:\\ 

\begin{equation}
 \left( \frac{\D}{\D t} + \uvec \cdot \nabla\right) \uvec = -\frac{1}{\rho_o}\nabla p +\frac{1}{4 \pi \rho_o \mu_o}(\bv{B} \cdot \nabla)\bv{B} + \nu \Delta \uvec 
\end{equation}

\begin{equation}
\frac{\partial \bv{B}}{\partial t}+ (\bv{u} \cdot \nabla) \bv{B}= (\bv{B} \cdot \nabla) \bv{u} + \eta \Delta \bv{B}
\end{equation}
\begin{equation}
\nabla\cdot\bv{u}=0,\qquad \nabla\cdot\bv{B}=0,
\end{equation}
in which appears the fluid density, $\rho_0$, the kinematic viscosity, $\nu$, and the magnetic diffusivity,
$\eta=\frac{1}{\mu_o \sigma}$ with $\mu_0$ the permeability and $\sigma$ the conductivity, and $\Delta=\nabla^2$ is the Laplacian.
Couette channel flow is assumed with distance
nondimensionalized by the half cross-stream channel width, L,  and velocity by the cross-stream boundary velocity, U.  The dimensionless equations are:

\begin{equation}
 \left( \frac{\D}{\D t} + \uvec \cdot \nabla\right) \uvec = -\nabla p +(\bv{B} \cdot \nabla)\bv{B} + \frac{1}{Re} \Delta \uvec 
\label{nondimNS}
\end{equation}

\begin{equation}
\frac{\partial \bv{B}}{\partial t}+ (\bv{u} \cdot \nabla) \bv{B}= (\bv{B} \cdot \nabla) \bv{u} + \frac{1}{Re_{m}}\Delta \bv{B}
\label{nondimind}
\end{equation}\\
in which appear the Reynolds number and magnetic Reynolds number defined in terms of the kinematic viscosity, $\nu$, and the magnetic diffusivity, $\eta$: 
\begin{equation}
Re=\frac{U L}{\nu}, ~~Re_m=\frac{U L}{\eta}.
\end{equation}

Periodic boundary conditions are imposed in the streamwise and spanwise directions and  impenetrable no-slip for the velocity field and vanishing of the magnetic field at the cross-stream boundaries. 
\begin{equation} \bv{u}(y=1)=1\hat{i}, \bv{B}(y=1)=0 
\end{equation}
\begin{equation} \bv{u}(y=-1)=-1\hat{i}, \bv{B}(y=-1)=0 
\end{equation}
The boundary condition for the magnetic field is the analogue of the no-slip condition on the velocity, so that the velocity and magnetic field components of the RSS satisfy structurally identical boundary conditions. It is a form of the pseudo-vacuum condition commonly used in dynamo simulations, which requires the tangential components of the magnetic field to vanish at the walls, $B_x=B_z=0$, with $\partial_y B_y=0$ following from $\nabla\cdot\bv{B}=0$. In addition, we set $B_y=0$ at the walls. With these boundary conditions, no wall-normal magnetic field is present at the walls, and therefore the dynamo mechanism identified in this work does not rely on a normal magnetic field generated at the boundaries.

\section{Formulation of MHD turbulence as a SSD}
Our focus is on studying composite velocity-magnetic field RSS dynamics using the MHD equations in SSD form.  The choice of Reynolds average is crucial in constructing an SSD that represents the underlying physics in the most transparent manner.  We choose the streamwise average for our Reynolds average.

Denoting an average in an independent variable by square brackets, the streamwise average becomes $[...]_x$ and using an overbar to compactly denote a streamwise averaged variable we write:\\
\begin{equation}
\overline{\bv{u}}=[\bv{u}]_x=\overline{u}_x(y,z,t) \hat{i}+ \overline{u}_y(y,z,t) \hat{j}+\overline{u}_z(y,z,t) \hat{k}
\end{equation}
\begin{equation}
\overline{\bv{B}}=[\bv{B}]_x=\overline{B}_x(y,z,t) \hat{i}+ \overline{B}_y(y,z,t) \hat{j}+\overline{B}_z(y,z,t) \hat{k}.
\end{equation}\\

We decompose state variables into streamwise mean and fluctuations:\\

\begin{equation}
\bv{u}(x,y,z,t)=\bv{\overline{u}}(y,z,t)+\bv{u'}(x,y,z,t)
\end{equation}
\begin{equation}
\bv{B}(x,y,z,t)=\overline{\bv{B}}(y,z,t)+\bv{B'}(x,y,z,t),
\end{equation}
\begin{equation}
p(x,y,z,t)=\overline{p}(y,z,t)+p'(x,y,z,t).
\end{equation}\\ 

The equations for the streamwise means are:\\

\begin{equation}
\frac{\partial \bv{\overline{u}}}{\partial t}=-(\bv{\overline{u}}\cdot \nabla)\bv{\overline{u}} - \overline{\bv{u'}\cdot \nabla \bv{u'}}   -\nabla \overline{p} +(\bv{\overline{B}} \cdot \nabla)\bv{\overline{B}} + \overline{(\bv{B'}\cdot \nabla)\bv{B'}} + \frac{1}{Re} \Delta \bv{\overline{u}}
\end{equation}
\begin{equation}
\frac{\partial \overline{\bv{B}}}{\partial t}= -(\overline{\bv{u}}\cdot \nabla)\overline{\bv{B}}+ (\overline{\bv{B}}\cdot \nabla) \overline{\bv{u}}- \overline{(\bv{u'}\cdot \nabla)\bv{B'}}+ (\overline{\bv{B'}\cdot \nabla)\bv{u'}}+\frac{1
}{Re_{m}} \Delta\overline{\bv{B}}.
\end{equation}\\

The fluctuation equations result from subtracting the mean equations from the full equations:\\

\begin{equation}
\frac{\partial \bv{u'}}{\partial t}=- (\bv{\overline{u}}\cdot \nabla)\bv{u'}- (\bv{u'}\cdot \nabla)\bv{\overline{u}}-\nabla p'+ (\overline{\bv{B}}\cdot \nabla)\bv{B'}+ (\bv{B'}\cdot \nabla)\overline{\bv{B}}+ \frac{1}{Re} \Delta \bv{u'}+\epsilon^{1/2} \bv{F_{u'}}\bv{\xi}(t)
\label{uprimeequation}
\end{equation}
\begin{equation}
\frac{\partial \bv{B'}}{\partial t}=-(\overline{\bv{u}}\cdot \nabla)\bv{B'}-(\bv{u'}\cdot \nabla) \overline{\bv{B}}+(\overline{\bv{B}}\cdot \nabla) \bv{u'}+ (\bv{B'}\cdot \nabla)\overline{\bv{u}}+\frac{1}{Re_{m}} \Delta \bv{B'}+ \epsilon^{1/2} \bv{F_{B'}}\bv{\xi}(t)
\label{Bprimeequation}
\end{equation}\\
in which a stochastic parameterization formed by the fluctuation structure matrices $\bv{F_{u'}},\bv{F_{B'}}$, a vector of independent temporally delta-correlated stochastic processes $\bv{\xi}(t)$, and an overall excitation amplitude $\epsilon$ replaces the non-linear fluctuation-fluctuation interaction terms in the momentum and induction equations \eqref{uprimeequation}, \eqref{Bprimeequation}. These stochastic terms also serve to parameterize any exogenous excitation in the model.  Note that with this  parameterization for the nonlinear terms in the fluctuation equations, the fluctuation equations become linear with respect to $\bv{u'}$ and $\bv{B'}$, and the resulting overall dynamics is quasi-linear.\\

We define the mean state variable as\\
\begin{equation}
\bv{\Gamma}=[\bv{\overline{u}}, \overline{\bv{B}}]^{T}.
\end{equation}\\
The mean equation can be written as:\\ 

\begin{equation}
\frac{\partial \bv{\Gamma}}{\partial t}=\begin{bmatrix}-(\bv{\overline{u}}\cdot \nabla)\bv{\overline{u}}-\nabla \overline{p}+ (\overline{\bv{B}}\cdot \nabla )\overline{\bv{B}}+\frac{1}{Re}\Delta \bv{\overline{u}} \\ -(\overline{\bv{u}}\cdot \nabla)\overline{\bv{B}}+ (\overline{\bv{B}}\cdot \nabla) \overline{\bv{u}}+\frac{1}{Re_{m}}\Delta\overline{\bv{B}}
\end{bmatrix}+\begin{bmatrix} -(\overline{\bv{u'}\cdot \nabla) \bv{u
'}}+ \overline{(\bv{B'}\cdot \nabla)\bv{B'}} \\ - \overline{(\bv{u'}\cdot \nabla)\bv{B'}}+ (\overline{\bv{B'}\cdot \nabla)\bv{u'}}
\end{bmatrix}
\end{equation}\\

We define the fluctuation state variable as:

\begin{equation}\bv{\phi}\equiv[ \bv{u'}, \bv{B'}]^{T}.\end{equation}\\
To formulate a SSD closed at second order, an equation for the infinite ensemble average fluctuation covariance $\bv{C}\equiv<\bv{\phi} \bv{\phi}^T>$ is required.
Under the assumption that the stochastic closure is white in time, an evolution equation for the infinite ensemble average covariance of the fluctuation state
can be obtained in the form of a time dependent Lyapunov equation \citep{Farrell 2019}:  

\begin{equation}
\partial_t \bv{C}=\bv{A}(\bv{\Gamma}) \bv{C} +\bv{C} \bv{A}(\bv{\Gamma})^{\dagger}+ \epsilon\bv{Q}
\end{equation}\\
in which $\bv{A}$ is the linear operator of the fluctuation equations \eqref{uprimeequation} and  \eqref{Bprimeequation}\\
and the stochastic closure is\\
\begin{equation} 
 \bv{Q}= \begin{bmatrix}\bv{Q}_{\bv{u'}\bv{u'}} & \bv{Q}_{\bv{u'}\bv{B'}} \\ \bv{Q}_{\bv{B'}\bv{u'}} & \bv{Q}_{\bv{B'}\bv{B'}}
 \end{bmatrix}
\end{equation}\\
in which the fluctuation excitation structure correlation matrices appear: 
\begin{equation}
\bv{Q_{u'u'}}=\bv{F_{u'}}\bv{F_{u'}}^{\dagger}, ~~ \bv{Q_{u'B'}}=\bv{F_{u'}}\bv{F_{B'}}^{\dagger}
\end{equation}
\begin{equation}
\bv{Q_{B'u'}}=\bv{F_{B'}}\bv{F_{u'}}^{\dagger}=\bv{Q_{u'B'}}^{\dagger}, ~~ \bv{Q_{B'B'}}=\bv{F_{B'}}\bv{F_{B'}}^{\dagger}.
\end{equation}\\
The diagonal blocks, $\bv{Q}_{\bv{u'u'}}$ and $\bv{Q}_{\bv{B'B'}}$, are the spatial covariances of the velocity and magnetic field excitation, while the off-diagonal blocks, $\bv{Q}_{\bv{u'B'}}$ and $\bv{Q}_{\bv{B'u'}}$, are the cross-covariances between the velocity and magnetic field excitation.
In this work the diagonal blocks are chosen to excite each degree of
freedom equally in  energy, which is accomplished by choosing them to be proportional to the identity.  In this paper the velocity and magnetic field fluctuations are excited equally, with a single amplitude $\epsilon$, and the diagonal blocks of $\bv{Q}$ are the normalized structures $\hat{\bv{Q}}_{\bv{u'u'}}$ and $\hat{\bv{Q}}_{\bv{B'B'}}$, whose normalization is specified in section \ref{compositeinstabilityresultssection}. \\

We refer to this second order SSD as the S3T SSD, where S3T stands for stochastic structural stability theory \citep{Farrell-Ioannou-2003, Farrell-Ioannou-2012}, and write it in compact form as:\\ 

\begin{equation}
\partial_t \bv{\Gamma}=\bv{G}(\bv{\Gamma})+\bv{L}(\bv{C})
\label{compositemeaneq}
\end{equation}
\begin{equation}
\partial_t \bv{C}=\bv{A}(\bv{\Gamma}) \bv{C} +\bv{C} \bv{A}(\bv{\Gamma})^{\dagger}+ \epsilon\bv{Q}
\label{compositecovarianceeq}
\end{equation}\\

where\\

\begin{equation}
\bv{G}(\bv{\Gamma})=\begin{bmatrix}-(\bv{\overline{u}}\cdot \nabla)\bv{\overline{u}}-\nabla \overline{p}+ (\overline{\bv{B}}\cdot \nabla )\overline{\bv{B}}+\frac{1}{Re}\Delta \bv{\overline{u}} \\ -(\overline{\bv{u}}\cdot \nabla)\overline{\bv{B}}+ (\overline{\bv{B}}\cdot \nabla) \overline{\bv{u}}+\frac{1}{Re_{m}}\Delta\overline{\bv{B}}
\end{bmatrix}
\end{equation}
\begin{equation}
\bv{L}(\bv{C})=\begin{bmatrix} -(\overline{\bv{u'}\cdot \nabla) \bv{u
'}}+ \overline{(\bv{B'}\cdot \nabla)\bv{B'}} \\ - \overline{(\bv{u'}\cdot \nabla)\bv{B'}}+ (\overline{\bv{B'}\cdot \nabla)\bv{u'}}
\end{bmatrix}.
\label{LCdef}
\end{equation}
The term $\bv{L}(\bv{C})$ is the fluctuation-fluctuation feedback on the mean state, which contains the Reynolds and Maxwell stresses and the turbulent electromotive force.
\begin{equation} 
 \bv{Q}= \begin{bmatrix}\bv{Q}_{\bv{u'}\bv{u'}} & \bv{Q}_{\bv{u'}\bv{B'}} \\ \bv{Q}_{\bv{B'}\bv{u'}} & \bv{Q}_{\bv{B'}\bv{B'}}
 \end{bmatrix}
\end{equation}\\

For the purpose of integration and to isolate as diagnostic variables the streamwise mean streamwise velocity and magnetic field components and the streamwise mean spanwise-cross stream velocity and magnetic field we take advantage of the velocity and magnetic fields being nondivergent to write the S3T SSD mean equations using as state variables the cross-stream-spanwise components of velocity and magnetic field expressed using  streamfunctions
 as 
$\bv{\Gamma}=[\overline{u}_x, \Psi, \overline{B}_x, \Xi]^{T}$ with the mean velocity and magnetic field streamfunctions respectively defined as:

\begin{equation}
-\partial_z\Psi= \overline{u}_y, ~~\partial_y \Psi= \overline{u}_z
\end{equation}
\begin{equation}
-\partial_z\Xi= \overline{B}_y, ~~\partial_y \Xi= \overline{B}_z.
\end{equation}\\

The mean state dynamics exclusive of the fluctuation-fluctuation term are:\\ 

\begin{equation}
\bv{G}(\bv{\Gamma})=\begin{bmatrix} G_{\overline{u}_x}(\bv{\Gamma})\\ G_{\Psi}(\bv{\Gamma}) \\ G_{\overline{B}_x}(\bv{\Gamma}) \\ G_{\Xi}(\bv{\Gamma})
\end{bmatrix},\end{equation}\\
in which are defined separately the equations for the mean components of the velocity and magnetic field:\\

\begin{equation} G_{\overline{u}_x}(\bv{\Gamma})= \partial_y\overline{u}_x \partial_z \Psi -\partial_z \overline{u}_x\partial_y \Psi-\partial_y \overline{B}_x \partial_z \Xi+\partial_z\overline{B}_x\partial_y\Xi+\frac{1}{Re}\Delta \overline{u}_x \end{equation}
\begin{equation}G_{\Psi}(\bv{\Gamma})= \Delta^{-1}[(\partial_{yy}-\partial_{zz})(\partial_y \Psi \partial_z \Psi-\partial_y\Xi\partial_z\Xi)-\partial_{yz}((\partial_y \Psi)^2-(\partial_z \Psi)^2-(\partial_y \Xi)^2+(\partial_z \Xi)^2)+\frac{1}{Re} \Delta \Delta \overline{\Psi}] 
\end{equation}
\begin{equation}
G_{\overline{B}_x}(\bv{\Gamma})=
\frac{\partial (-\overline{u}_x\partial_z\Xi+\partial_z\Psi \overline{B}_x)}{\partial y}-\frac{\partial (\partial_y \Psi \overline{B}_x-\overline{u}_x \partial_y \Xi)}{\partial z}+\frac{1}{Re_{m}}\Delta \overline{B}_x
\end{equation}
\begin{equation}
G_{\Xi}(\bv{\Gamma})=
\Delta^{-1}[-(\partial_{yy}+\partial_{zz})(-\partial_{z}\Psi\partial_y \Xi+\partial_y \Psi \partial_z \Xi)+ \frac{1}{Re_m}\Delta \Delta \Xi]
\end{equation}\\

The fluctuation-fluctuation term in the mean equation \eqref{LCdef}, which contains the Reynolds and Maxwell stresses and the turbulent electromotive force, becomes:\\

\begin{equation}
\bv{L}(\bv{C})=\begin{bmatrix}-\partial_y \overline{u_x'u_y'}-\partial_z \overline{u_x'u_z'}+\partial_y \overline{B_x'B_y'}+\partial_z \overline{B_x'B_z'} \\ \Delta^{-1} [(\partial_{yy}-\partial_{zz})(-\overline{u_y'u_z'}+\overline{B_y'B_z'})-\partial_{yz}(\overline{u_z'^2}-\overline{u_y'^2}- \overline{B'^2_z}+\overline{B'^2_y})]\\
+\frac{\partial (\overline{u_x'B'_y-u_y'B_x'})}{\partial y}-\frac{\partial (\overline{u_z'B_x'-u_x'B_z'})}{\partial z}\\
\Delta^{-1}[-(\partial_{yy}+\partial_{zz})(\overline{u_y'B_z'- u_z' B_y'})]
\end{bmatrix}
\label{LRSCKeq}
\end{equation}\\

.

In a similar manner, for the purpose of integration the fluctuation covariance equation \eqref{compositecovarianceeq} is written in terms of cross stream components of velocity and magnetic field and cross stream components of curl of velocity and of magnetic field. 
To this end, the
fluctuation state variable is written as 
$\bv{\phi}=[u_y', \omega_y', B_y', \zeta_y']^{T}$ with the velocity and magnetic curl quantities:\\
 
\begin{equation}\omega_y'=\partial_z u_x'- \partial_x u_z'\end{equation}
\begin{equation} \zeta'_{y}= \partial_z B_x'-\partial_x B_z'.
\end{equation}\\

The ensemble average fluctuation covariance  $\bv{C}\equiv<\bv{\phi \phi^{T}}>$ is expressed using these variables and the blocks of
$\bv{Q}$ are  consistently transformed \citep{Kim Langmuir, Farrell-Ioannou-2012}.  The Lyapunov equation \eqref{compositecovarianceeq}
is then integrated in these variables.

\section{S3T stability formulation}
Assuming an equilibrium S3T state, $(\bv{\Gamma}_e, \bv{C}_e)$, in which the time derivatives in \eqref{compositemeaneq} and \eqref{compositecovarianceeq} vanish:\\
\begin{equation}
\bv{\Gamma}_e=\begin{bmatrix} \overline{u}_{xe} \\ \Psi_e \\ \overline{B}_{xe} \\ \Xi_e
\end{bmatrix}
\end{equation}\\
and following \citep{Farrell-Ioannou-2012,Farrell 2019,Farrell-Ioannou-2016-bifur} perturbation equations linearized around the SSD equilibrium state  ($\bv{\Gamma}_e$, $\bv{C}_e$) can be obtained:
\begin{equation}\partial_t(\delta \bv{\Gamma})=\sum_{i}^{} \frac{\partial \bv{G}}{\partial \Gamma_i}|_{\Gamma_{e}} \delta \Gamma_i+ \bv{L}(\delta \bv{C})
\label{meanS3Tlinear}
\end{equation}
\begin{equation}\partial_t(\delta \bv{C})=\bv{A}(\bv{\Gamma}_{e})\delta \bv{C}+ \delta \bv{C} \bv{A}(\bv{\Gamma}_{e})^{\dagger}+ \delta \bv{A} \bv{C}_{e}+\bv{C}_{e}\delta \bv{A}^{\dagger}\label{covarianceS3Tlinear}\end{equation}
where 
\begin{equation}\delta \bv{A}=\bv{A}(\bv{\Gamma}_e+ \delta \bv{\Gamma})-\bv{A}(\bv{\Gamma_e}).
\end{equation}\\

Equations \eqref{meanS3Tlinear} and \eqref{covarianceS3Tlinear} comprise the formulation of the linear perturbation S3T dynamics.\\

\section{Stability analysis of the composite velocity and magnetic field RSS in the S3T SSD  of turbulent Couette flow}
\label{compositeinstabilityresultssection}
In this section, we focus on the linear stability of \eqref{meanS3Tlinear} - \eqref{covarianceS3Tlinear}.  While in non-conducting Couette flow the S3T SSD supports velocity state instabilities with RSS structure \citep{Farrell-Ioannou-2012, Farrell-Ioannou-2016-bifur},   we find that both the velocity field and the magnetic field participate synergistically in destabilizing the RSS in turbulent conducting Couette flow.  

 We  use Fourier series to represent fluctuations in the streamwise direction, x, and the spanwise direction, z. The channel domain sizes in $x$ and $z$ are $L_x=1.75 \pi$ and $L_z=1.2\pi$, respectively.  A single streamwise wavenumber and $8$ spanwise wavenumbers are retained in x and z along with $N_y=25$ grid points in $y$.
 Retaining a single wavenumber in x is consistent with self-sustaining turbulence in our model being supported solely by this wavenumber \citep{ Farrell-Ioannou-2012}. The normalized excitation structures $\hat{\bv{Q}}_{\bv{u'u'}}$ and $\hat{\bv{Q}}_{\bv{B'B'}}$ are chosen such that $\epsilon=1$ results in volume averaged RMS
fluctuation velocity  and magnetic field being 1 $\%$ of the maximum velocity of the Couette profile. Explicitly, $[\sqrt{\overline{u_x'^2+u_y'^2+u_z'^2}}]_{y,z}=0.01$ and $[\sqrt{\overline{B_x'^2+B_y'^2+B_z'^2}}]_{y,z}=0.01$. Results are presented for $Re=Re_m=400$.

\begin{figure}
\centering{
\begin{subfigure}{\textwidth} 
\centering{\caption{}
\makebox[\linewidth][l]{\hspace{0.1836\linewidth}\includegraphics[width=0.4639\linewidth]{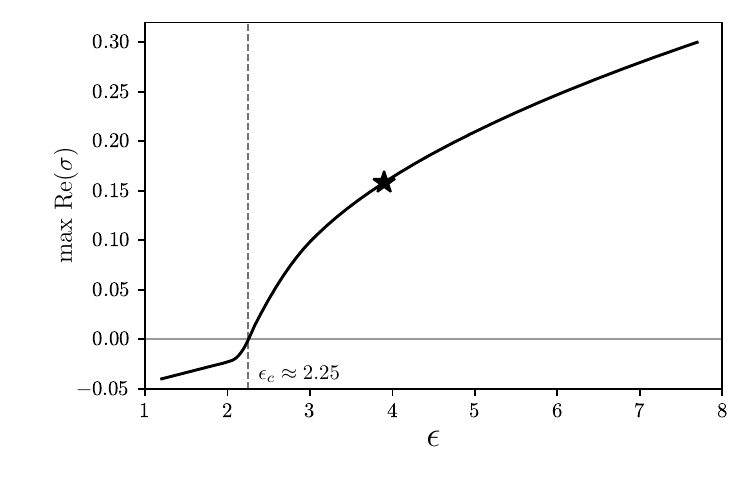}}
\label{fig:1a}} 
\end{subfigure}
\begin{subfigure}{\textwidth}\centering{ \caption{}
\makebox[\linewidth][l]{\hspace{0.1972\linewidth}\includegraphics[width=0.5310\linewidth]{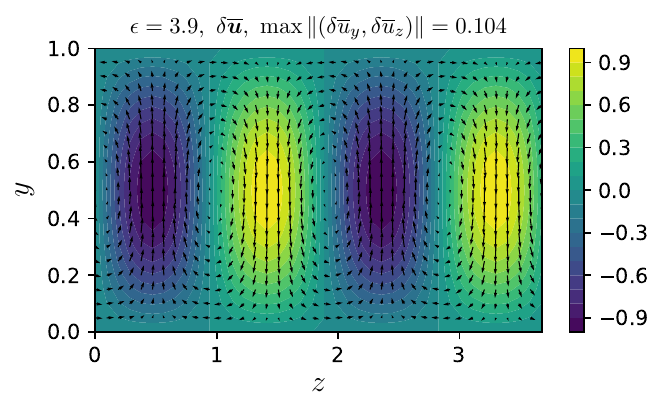}} \label{fig:1b}} 
\end{subfigure}
\begin{subfigure}{\textwidth} 
\centering{
\caption{}
\makebox[\linewidth][l]{\hspace{0.1972\linewidth}\includegraphics[width=0.5418\linewidth]{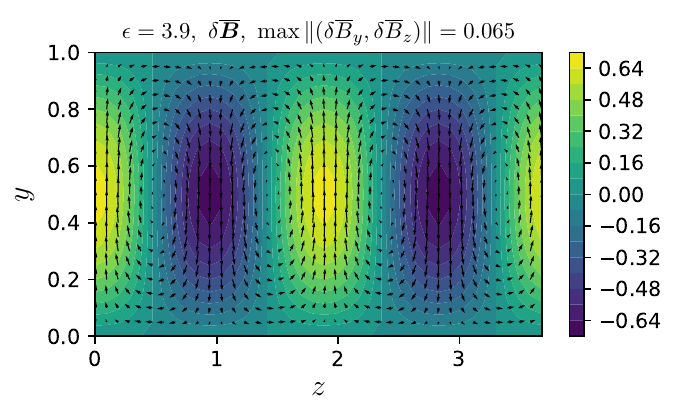}} \label{fig:1c}} 
\end{subfigure}
}
\caption{Panel(a): S3T stability diagram for $\epsilon$ at $Re=Re_m=400$, showing the growth rate of the most unstable RSS perturbation, $\max\,\mathrm{Re}(\sigma)$, as a function of $\epsilon$, where $\sigma$ is the eigenvalue of the linearized S3T system \eqref{meanS3Tlinear}--\eqref{covarianceS3Tlinear}, with eigenmodes of the form $(\delta\bv{\Gamma},\delta\bv{C})\,e^{\sigma t}$. The S3T SSD equilibrium is unstable to RSS perturbations for $\epsilon\gtrsim 2.25$. Panel (b): streamwise velocity contours and spanwise/cross-stream velocity vectors of most unstable eigenmode. Panel (c): streamwise magnetic field  contours and spanwise/cross-stream magnetic field vectors of most unstable eigenmode. Eigenfunction structures have been normalized so that the maximum value of $\delta \overline{u}_x$ is unity. 
The eigenmode shown in panels (b) and (c) is at $\epsilon=3.9$, marked by a star in panel (a).
 \label{fig:compositeinstability}}
\end{figure}

The S3T stability diagram is shown in panel a of figure \ref{fig:compositeinstability}, in which the growth rate of the most unstable RSS perturbation, $\max\,\mathrm{Re}(\sigma)$, is shown as a function of $\epsilon$. The S3T SSD equilibrium becomes unstable to RSS perturbations for $\epsilon\gtrsim 2.25$, beyond which the growth rate increases monotonically with $\epsilon$. 
Panels b and c of this figure show the velocity field structure $\delta{\overline{\bv{u}}}$ and magnetic field structure $\delta{\overline{\bv{B}}}$ of the unstable eigenmode at $\epsilon=3.9$.  Noteworthy are that the velocity and magnetic field components of the RSS are generally similar in structure despite the physically distinct mechanisms involved in the NS and induction equations.  The amplitude of the fields being similar indicates approximate energy equipartition in view of the scaling  used for the velocity and magnetic fields.  The characteristic 90 degree phase shift between the $\delta \overline{\bv{u}}$ and $\delta \overline{\bv{B}}$ fields indicates a synergy operating between these fields in maintaining their composite structure.

\label{S3Tinstabiltysection}
\section{Equilibration of  RSS instabilities in the S3T SSD}
\begin{figure}
\centering{
\includegraphics[width=0.85\linewidth]{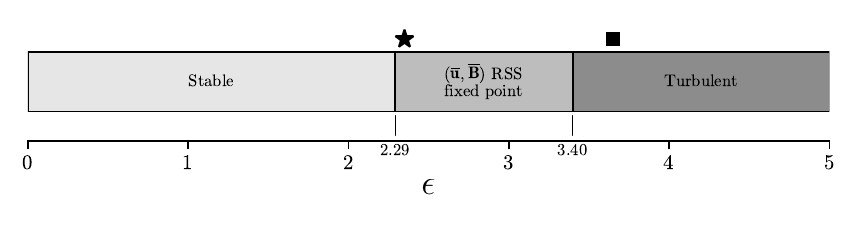}
}
\caption{Equilibrium regime diagram as a function of $\epsilon$ at $Re=Re_m=400$: the S3T SSD is stable for $\epsilon\lesssim 2.3$, equilibrates to a fixed point RSS supporting both $\overline{\bv{u}}$ and $\overline{\bv{B}}$ for $2.3\lesssim \epsilon\lesssim 3.4$, and transitions to turbulence for $\epsilon\gtrsim 3.4$. The regimes are assigned from the long-time behaviour of the S3T SSD integrated from the equilibrium perturbed by its most unstable eigenmode: ``stable'' when the perturbation decays, ``fixed point'' when the integration converges to a time-independent RSS, and ``turbulent'' when the state is chaotic in both space and time. The star and square mark the fixed point ($\epsilon=2.35$) and turbulent ($\epsilon=3.65$) cases examined below.} 
	\label{fig:equilibriumregime}
\end{figure}

The S3T SSD is a nonlinear system with an  equilibration mechanism consistent with its underlying  MHD dynamics.  Perturbing the Couette S3T SSD model
with unstable RSS eigenmodes results in establishing a finite-amplitude equilibrium consisting of
a fixed point RSS, or transition to a time-dependent  or turbulent state depending on
parameter values. As a summary of our findings and to set the stage for presenting our
results, an equilibrium regime diagram as a function of $\epsilon$ is shown in figure \ref{fig:equilibriumregime}. 
We turn now to examine salient aspects of these equilibrium regimes.

As the excitation, $\epsilon$, increases, the S3T SSD passes through three regimes: it is stable for $\epsilon\lesssim 2.3$, equilibrates to a fixed point RSS in which both a coherent mean velocity, $\overline{\bv{u}}$, and a coherent mean magnetic field, $\overline{\bv{B}}$, are maintained for $2.3\lesssim \epsilon\lesssim 3.4$, and becomes turbulent for $\epsilon\gtrsim 3.4$.
Unlike analogous regime diagrams for nonconducting wall bounded shear flow \citep{Farrell-Ioannou-2016-bifur}, when the velocity and magnetic field excitations differ the S3T SSD also supports a second, distinct fixed point regime in which the RSS has only a coherent mean velocity, $\overline{\bv{u}}$, component (not shown).
\begin{figure}
\centering{
\includegraphics[width=0.75\linewidth]{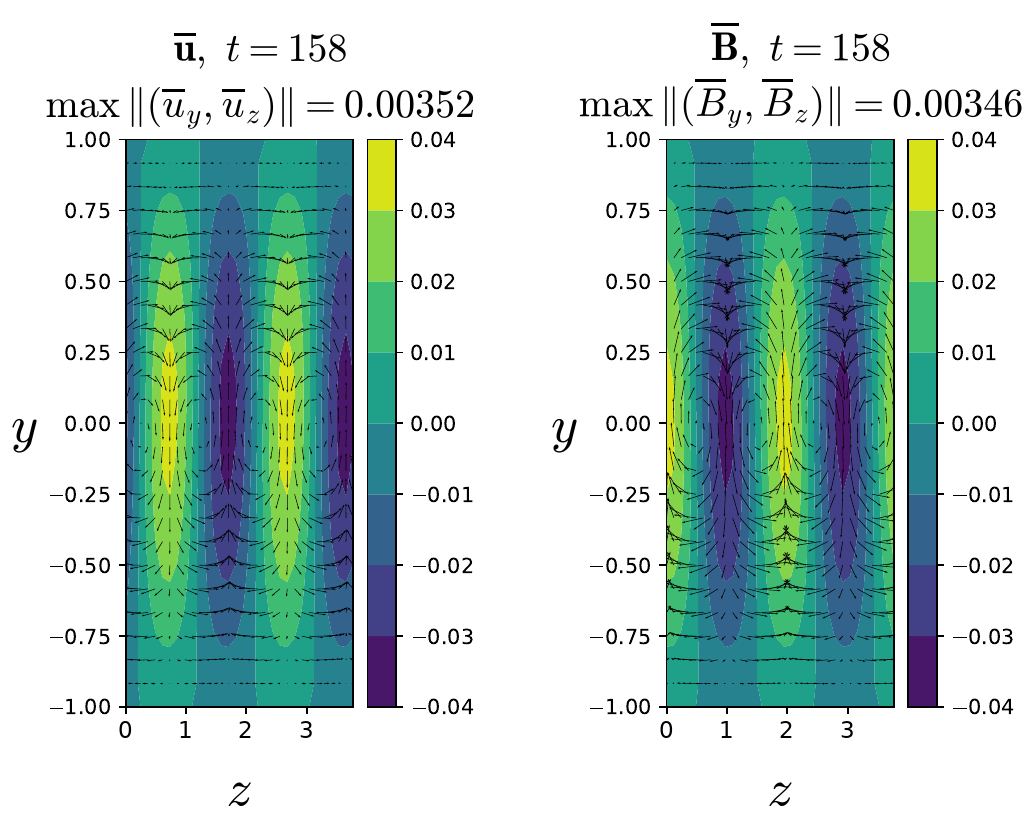}

}
    \caption{Fixed point equilibrium RSS supporting finite amplitude components in both the mean velocity field $\overline{\bv{u}}$ and the mean magnetic field $\overline{\bv{B}}$. On the left: streak velocity $\overline{u}_{xs}=\overline{u}_x-[\overline{u}_x]_z$ (contours) and roll velocity $(\overline{u}_{y},\overline{u}_z)$ (vectors). On the right: streak magnetic field $\overline{B}_{xs}=\overline{B}_x-[\overline{B}_x]_z$ (contours) and roll  magnetic field $(\overline{B}_y,\overline{B}_z)$ (vectors).  Parameters: $\epsilon=2.35$, $Re=Re_m=400$} 
	\label{fig:fixedpointsol}
\end{figure}
In figure \ref{fig:fixedpointsol} is shown the RSS structure of a fixed point equilibrium at $\epsilon=2.35$ (marked by a star in figure \ref{fig:equilibriumregime}) in which both a coherent velocity field and  a coherent magnetic field are supported.  In this fixed-point equilibrium RSS both the roll velocity field and the poloidal magnetic field components of the RSS, along with the $90$ degree phase relation between these components, agree with the structure of the S3T perturbation instability as shown in figure \ref{fig:compositeinstability}. We conclude that the unstable RSS mode qualitatively predicts the equilibrium structure of the finite amplitude equilibrium.  As shown in figure \ref{fig:equilibriumregime} above, a region of parameter space exists with fixed point equilibria in which the RSS maintains finite amplitude components in both the roll and streak velocity and the poloidal and toroidal magnetic field.

\begin{figure}
\subfloat[]{%
            \includegraphics[width=.48\linewidth]{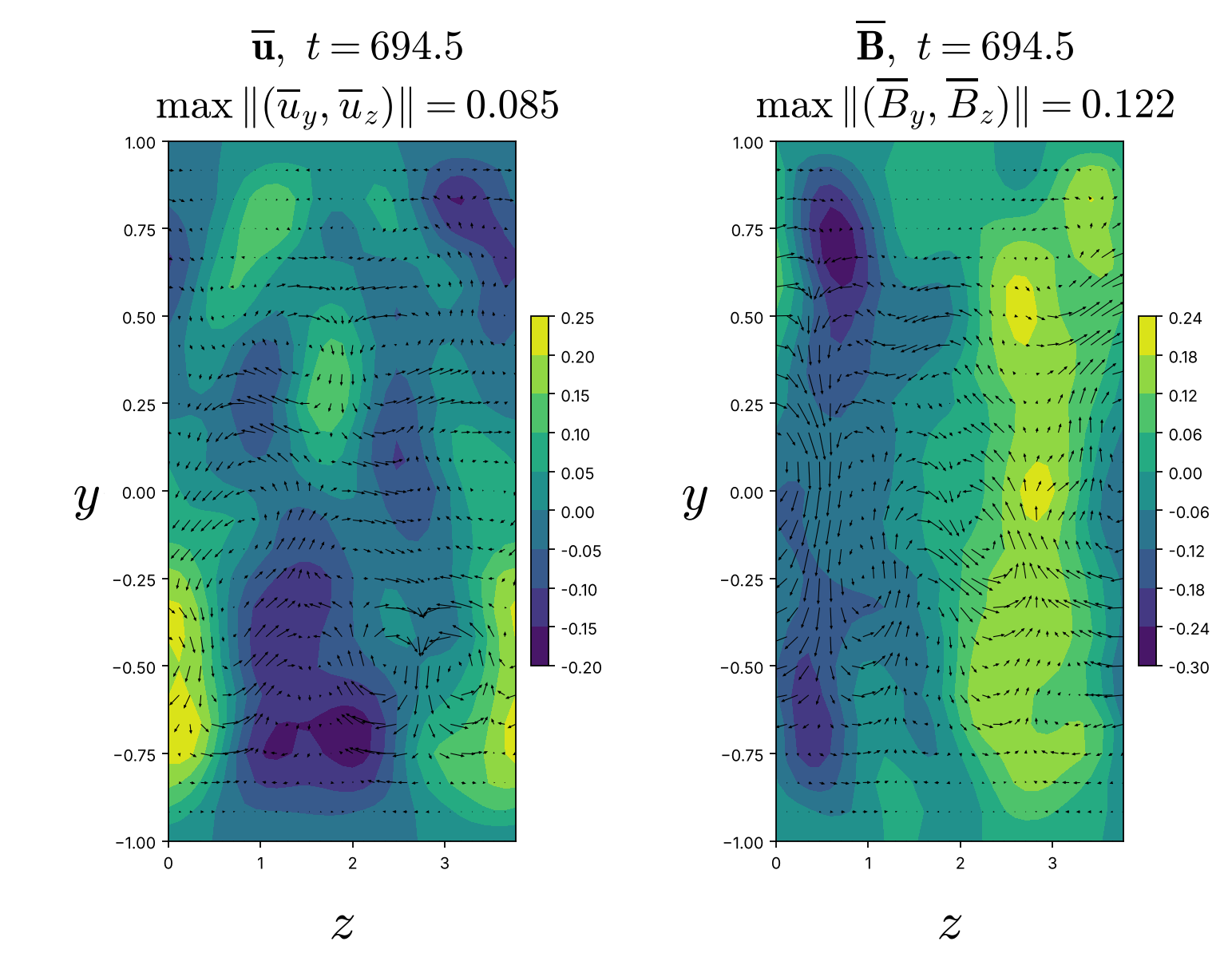}
            \label{subfig:a}%
        }\hfill
        \subfloat[]{%
            \includegraphics[width=.48\linewidth]{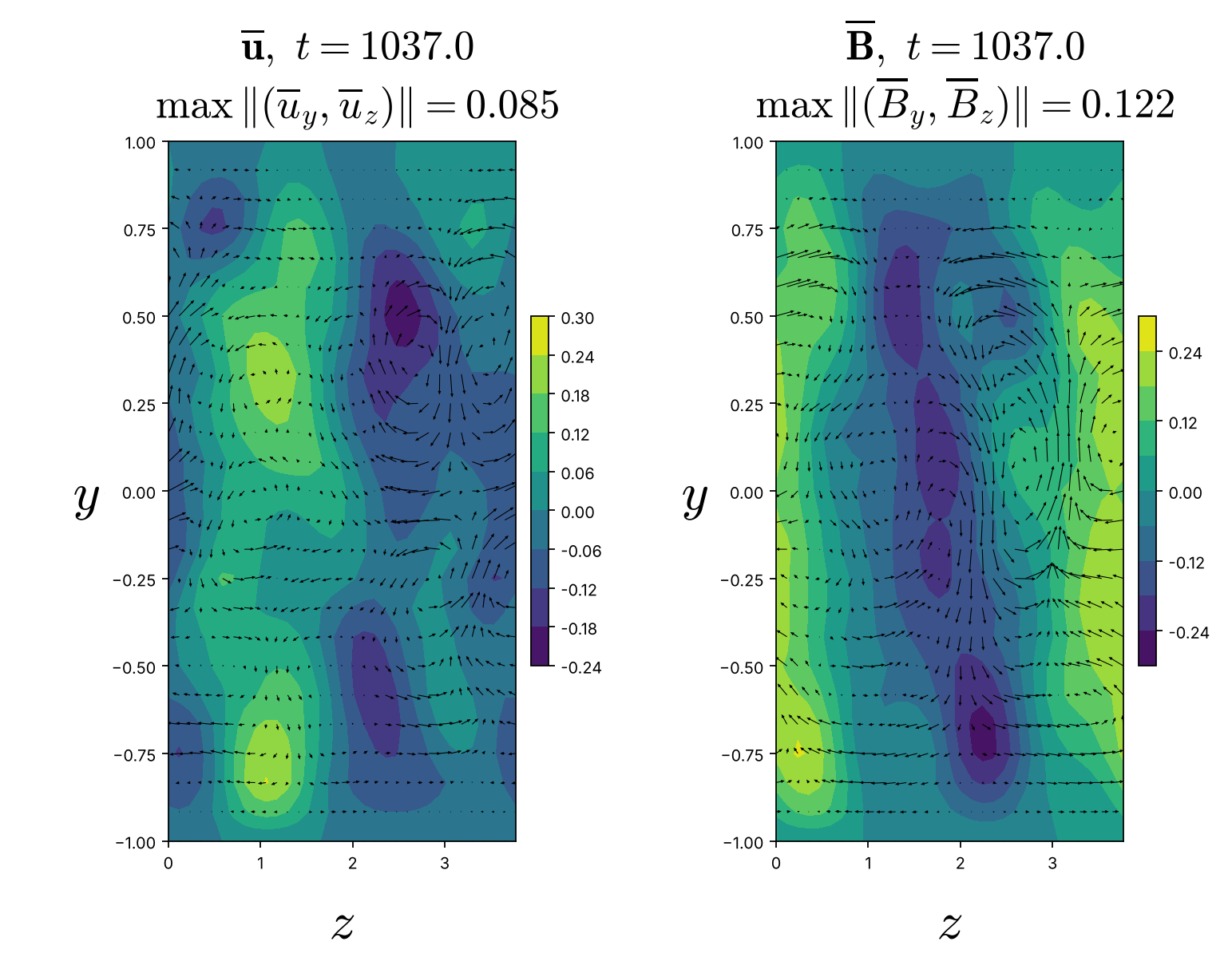}
            \label{subfig:b}%
        }\\
        \subfloat[]{%
            \includegraphics[width=.48\linewidth]{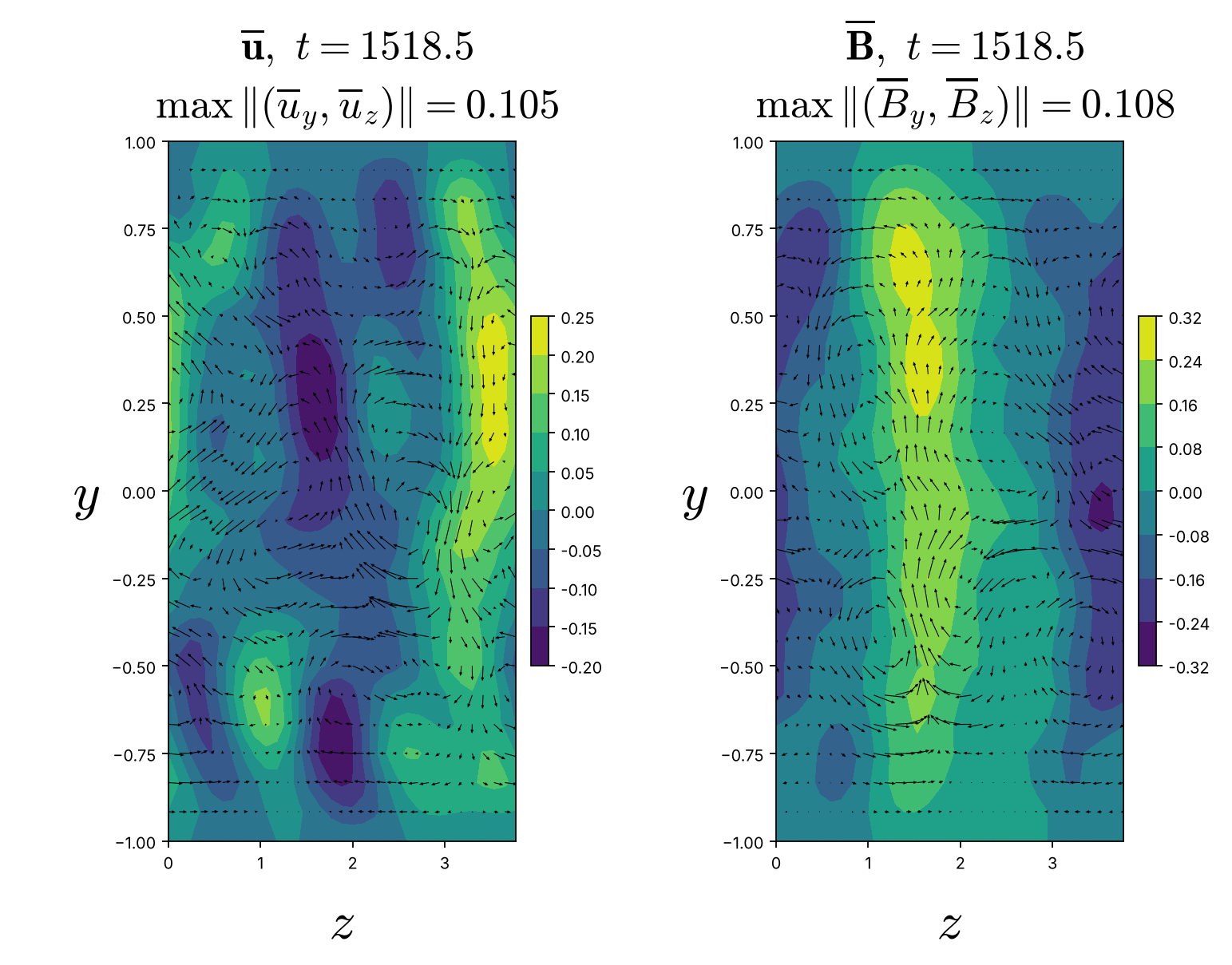}
            \label{subfig:c}%
        }\hfill
        \subfloat[]{%
            \includegraphics[width=.48\linewidth]{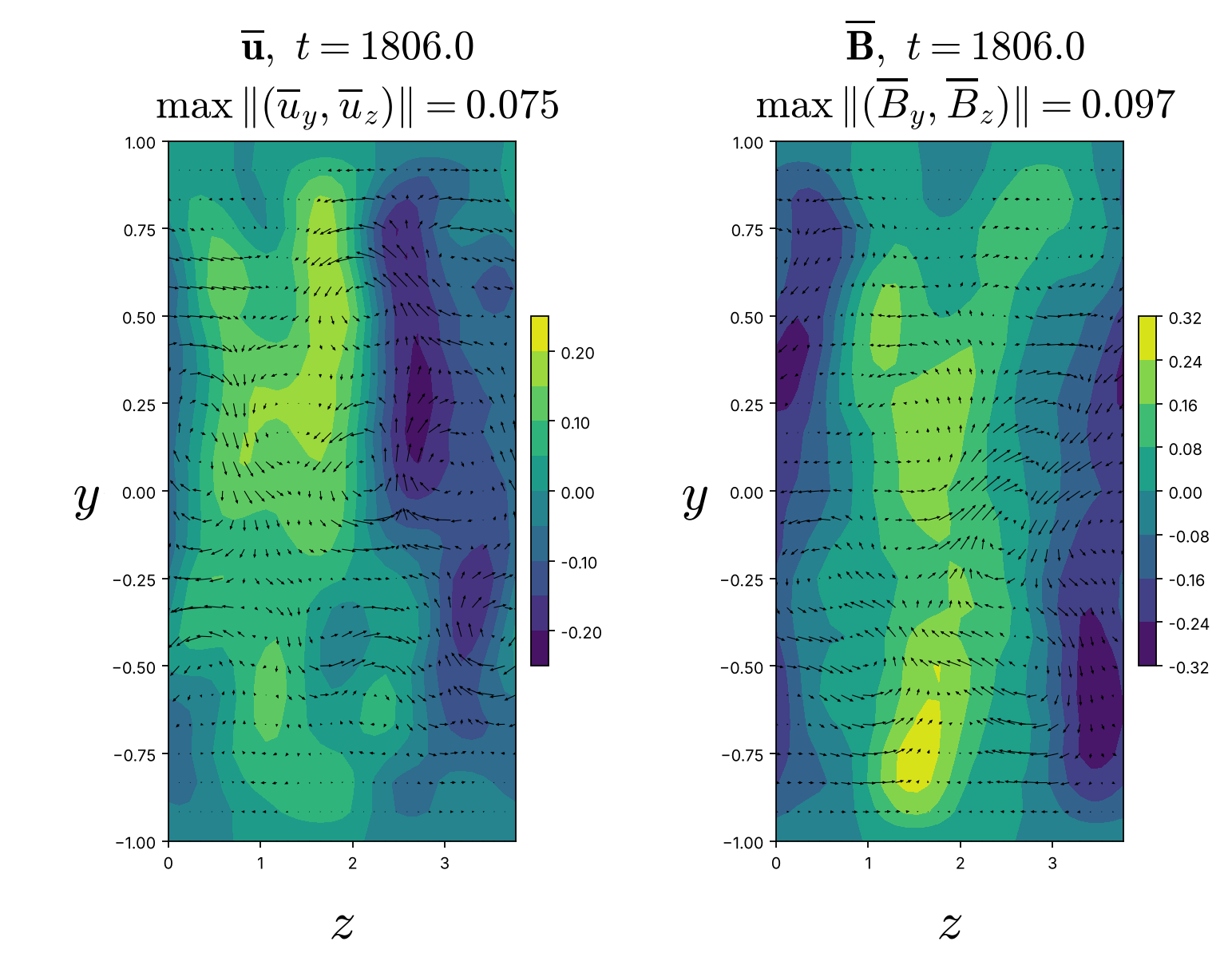}
            \label{subfig:d}%
        }
        
\caption{Snapshots of turbulent equilibria supporting finite amplitude components in both the mean velocity field $\overline{\bv{u}}$ and the mean magnetic field $\overline{\bv{B}}$. For each snapshot, on the left: streak velocity $\overline{u}_{xs}=\overline{u}_x-[\overline{u}_x]_z$ (contours) and roll velocity $(\overline{u}_{y},\overline{u}_z)$ (vectors). On the right: streak magnetic field $\overline{B}_{xs}=\overline{B}_x-[\overline{B}_x]_z$ (contours) and roll  magnetic field $(\overline{B}_y,\overline{B}_z)$ (vectors).  Parameters: $\epsilon=3.65$, $Re=Re_m=400$}
\label{fig:turbulentsnapshot}
\end{figure}

Snapshots of turbulent equilibria at $\epsilon=3.65$ (marked by a square in figure \ref{fig:equilibriumregime}) are shown in figure \ref{fig:turbulentsnapshot}.
 
\section{Self sustaining RSS turbulence}

In wall-bounded hydrodynamic turbulence a turbulent RSS state can self sustain with $\epsilon_{\bv{u'u'}}=0$ once it has transitioned to turbulence \citep{Farrell-Ioannou-2012}.
It is  of interest to  inquire whether the turbulent MHD RSS can sustain  without the presence of $\epsilon$. 
To address this question we excite a turbulent RSS state using $\epsilon$ and then set $\epsilon=0$ as a diagnostic  for the existence of a self sustaining turbulent state.

\begin{figure}
\subfloat[]{%
            \includegraphics[width=.48\linewidth]{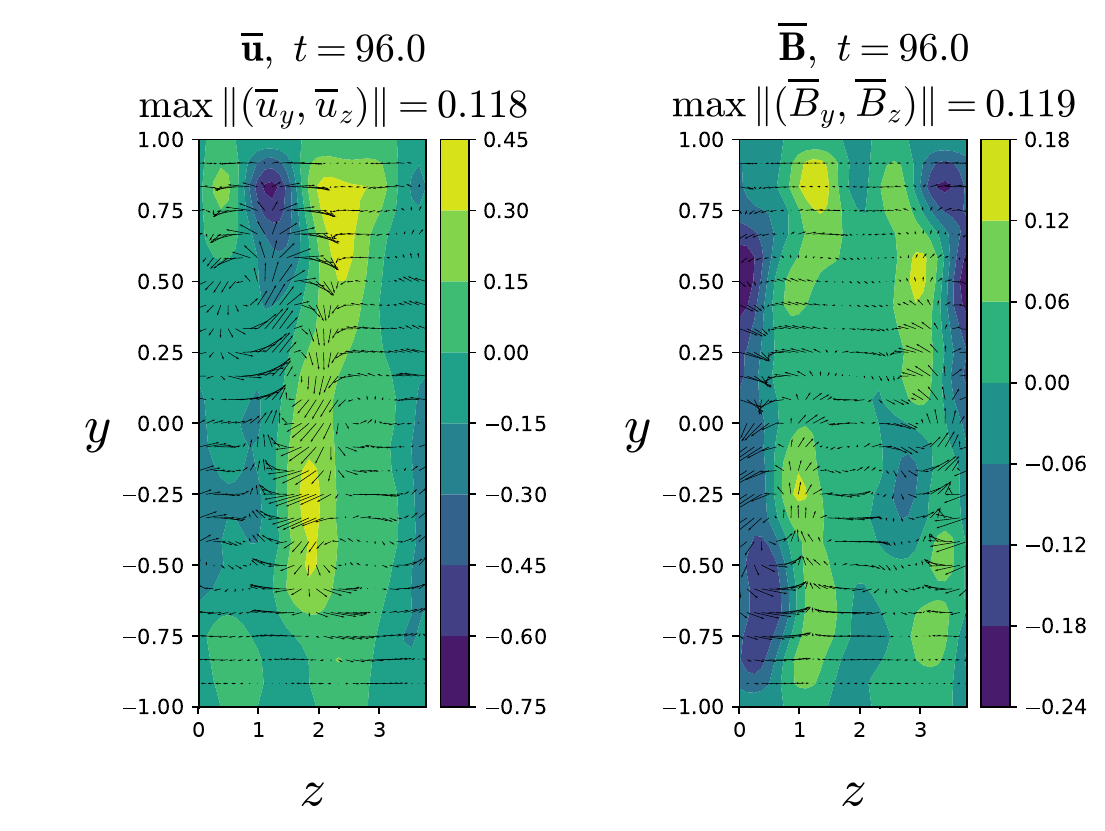}
            \label{subfig:a}%
        }\hfill
        \subfloat[]{%
            \includegraphics[width=.48\linewidth]{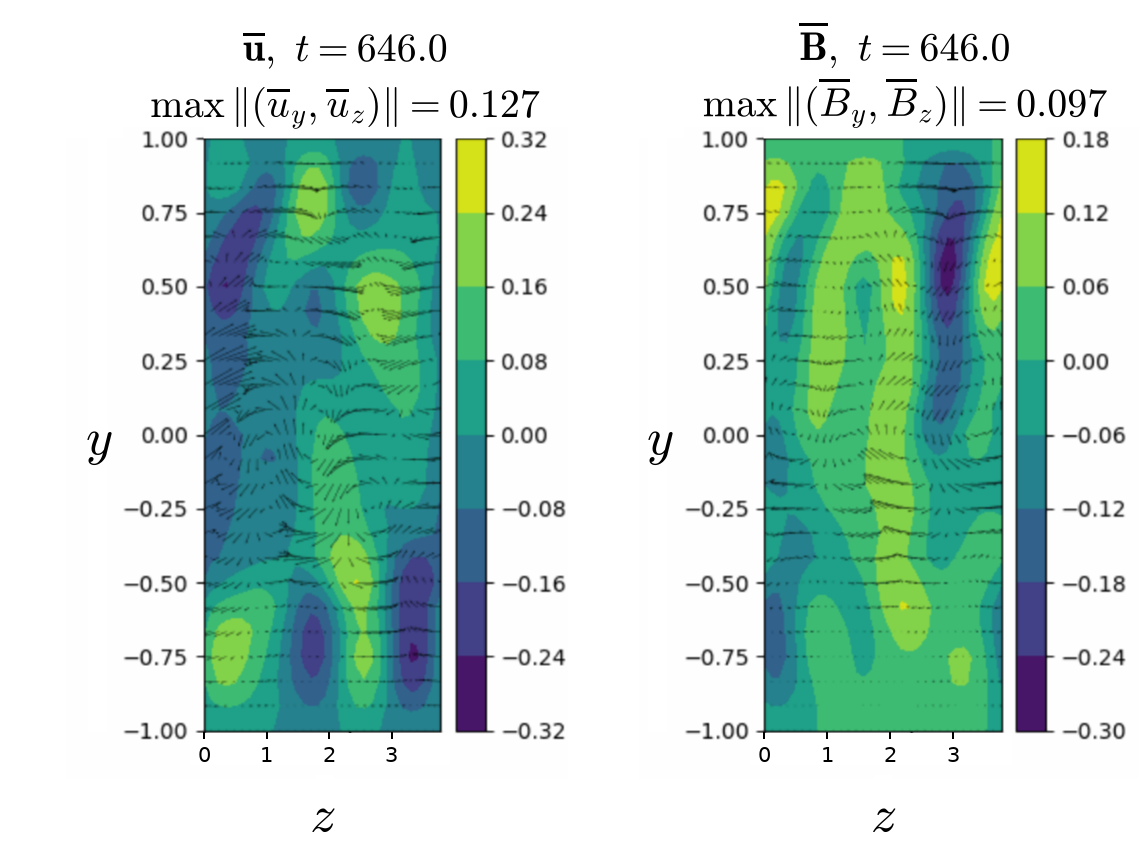}
            \label{subfig:b}%
        }\\
        \subfloat[]{%
            \includegraphics[width=.48\linewidth]{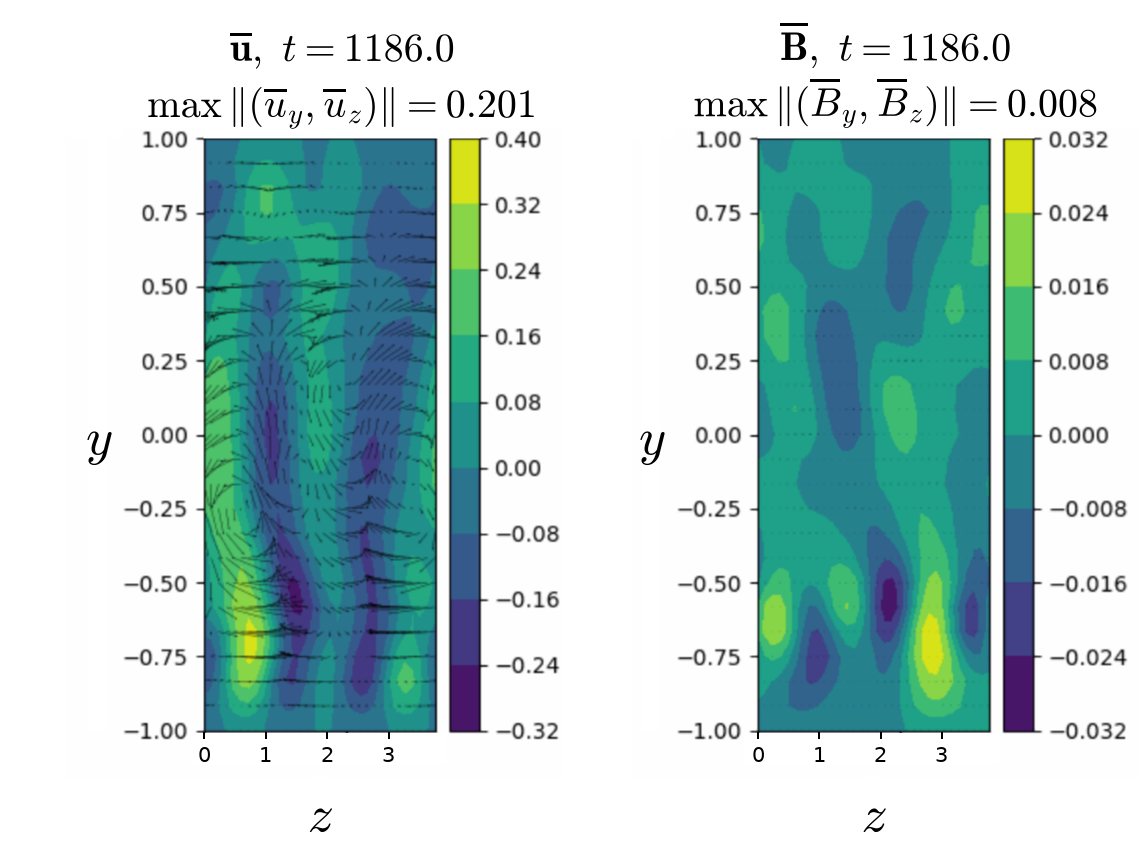}
            \label{subfig:c}%
        }\hfill
        \subfloat[]{%
            \includegraphics[width=.48\linewidth]{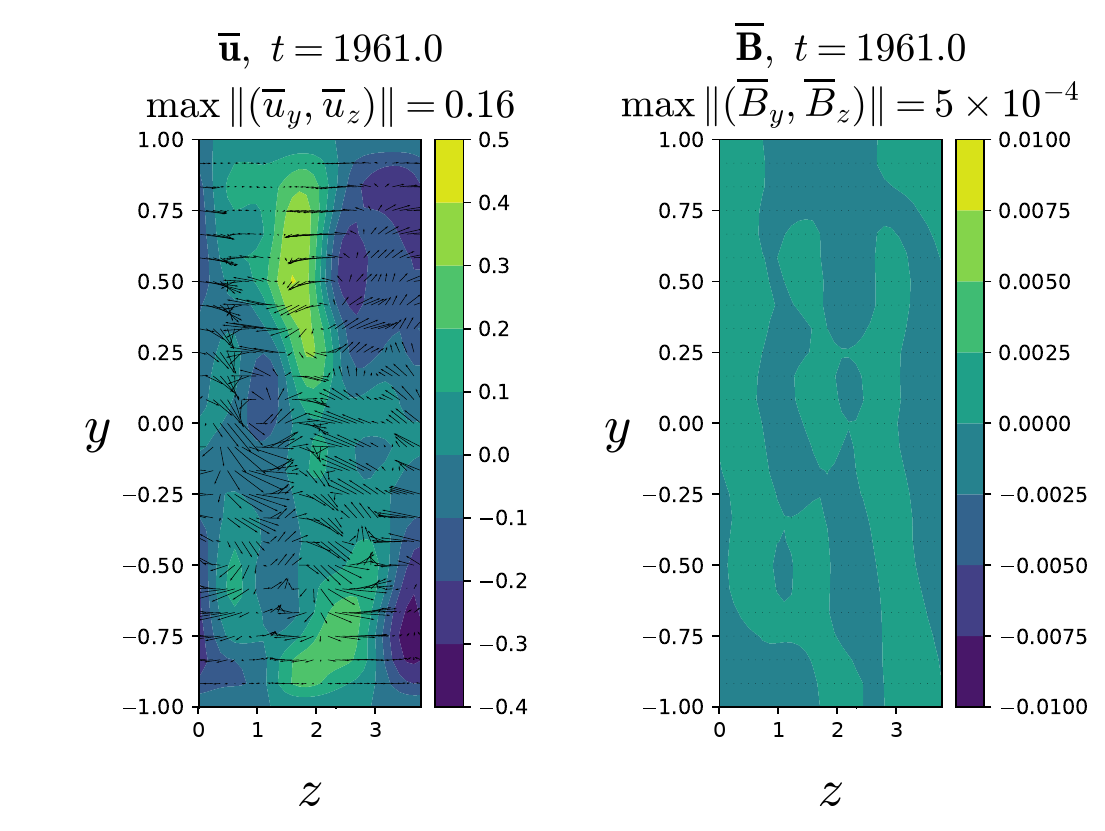}
            \label{subfig:d}%
        }
        
\caption{Snapshots of self sustaining turbulent equilibria sustaining finite amplitude components in only the mean velocity $\overline{\bv{u}}$. For each snapshot, on the left: streak velocity $\overline{u}_{xs}=\overline{u}_x-[\overline{u}_x]_z$ (contours) and roll velocity $(\overline{u}_{y},\overline{u}_z)$ (vectors). On the right: streak magnetic field $\overline{B}_{xs}=\overline{B}_x-[\overline{B}_x]_z$ (contours) and roll  magnetic field $(\overline{B}_y,\overline{B}_z)$ (vectors). (Note the rapid decrease with time in the amplitude of the magnetic field component); Parameters: $Re=Re_m=400$ ($Pr_m=1<Pr_{mc}\approx1.25$; cf.\ figure~\ref{fig:Prmbifur}), $\epsilon=0$.}
\label{fig:uonlySSP}
\end{figure}

\begin{figure}
\centering{
\includegraphics[width=0.75\linewidth]{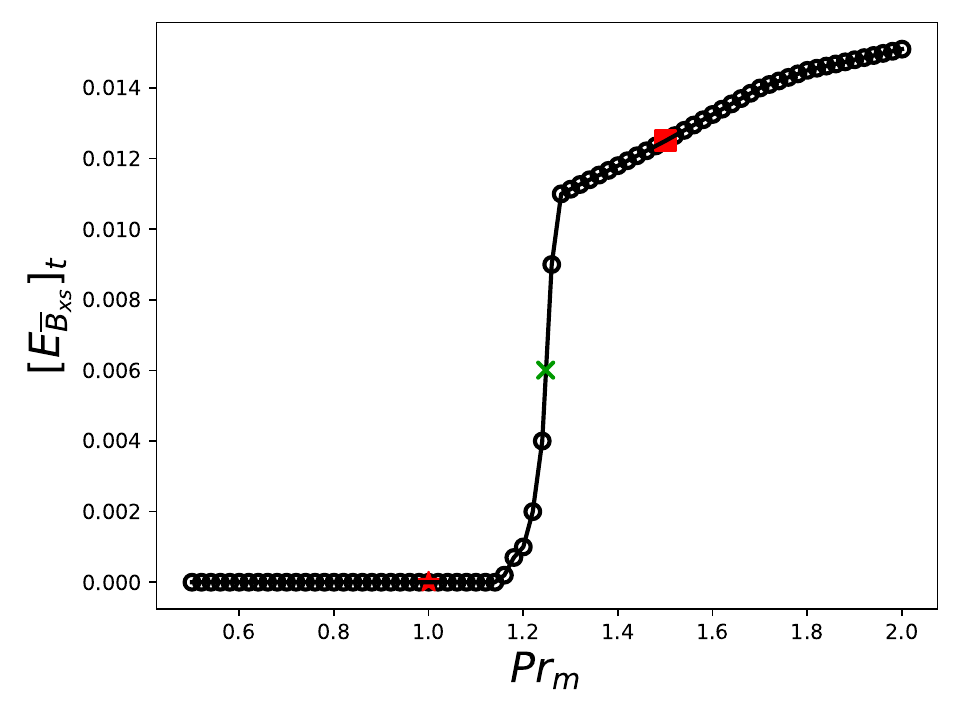}
}
\caption{Bifurcation to a maintained time-averaged streak magnetic energy, $[E_{\overline{B}_{xs}}]_t$, by the self-sustaining turbulent state as a function of magnetic Prandtl number, $Pr_m$. The critical Prandtl number at $Pr_{mc}\approx1.25$ is indicated by a green cross. An example of a state without a maintained field at $Pr_m=1.0$ (red star) is shown in figure~\ref{fig:uonlySSP} and a state with a maintained magnetic field at $Pr_m=1.5$ (red square) is shown in figure~\ref{fig:uBSSP}. The other parameters are $Re=400$ with $\epsilon=0$.}
	\label{fig:Prmbifur}
\end{figure}

More explicitly, after developing fully turbulent states, $\epsilon$ is set to zero resulting in the following equations:\\
\begin{equation}
\partial_t \bv{\Gamma}=\bv{G}(\bv{\Gamma})+ \bv{L}(\bv{C})
\label{meaneqSSP}
\end{equation}
\begin{equation}
\partial_t \bv{C}=\bv{A}(\bv{\Gamma}) \bv{C} +\bv{C} \bv{A}(\bv{\Gamma})^{\dagger}.
\label{coveqSSP}
\end{equation}\\
Shown in figure $\ref{fig:uonlySSP}$ is the result of eliminating the excitation in our example case at $Re=400, Pr_m=1$ (corresponding to a star in figure $\ref{fig:Prmbifur}$) showing that only the velocity self sustains while the magnetic field decays. 
At a given $Re$, existence of a  SSP supporting both a coherent velocity and a coherent magnetic field component in the RSS depends on  magnetic Prandtl number:\\
\begin{equation}
Pr_m=\frac{\nu}{\eta}=\frac{Re_{m}}{Re}.
\end{equation}\\
To demonstrate this, 
we show in figure \ref{fig:Prmbifur} the time averaged value of magnetic streak energy: 

\begin{equation}E_{\overline{B}_{xs}}\equiv\frac{1}{2}[\overline{B}_{xs}^2 ]_{y,z},
\label{EBxsdef}
\end{equation}\\
which is diagnostic of the existence of a coherent magnetic field component, as a function of  $Pr_{m}$.    Bifurcation from a  self sustaining state in the velocity field only to a self sustaining state in both the velocity and the magnetic field, which is indicative of a self sustaining dynamo, is seen in figure \ref{fig:Prmbifur} to occur at critical magnetic Prandtl number $Pr_{mc}\approx 1.25$.

\begin{figure}
\subfloat[]{%
            \includegraphics[width=.48\linewidth]{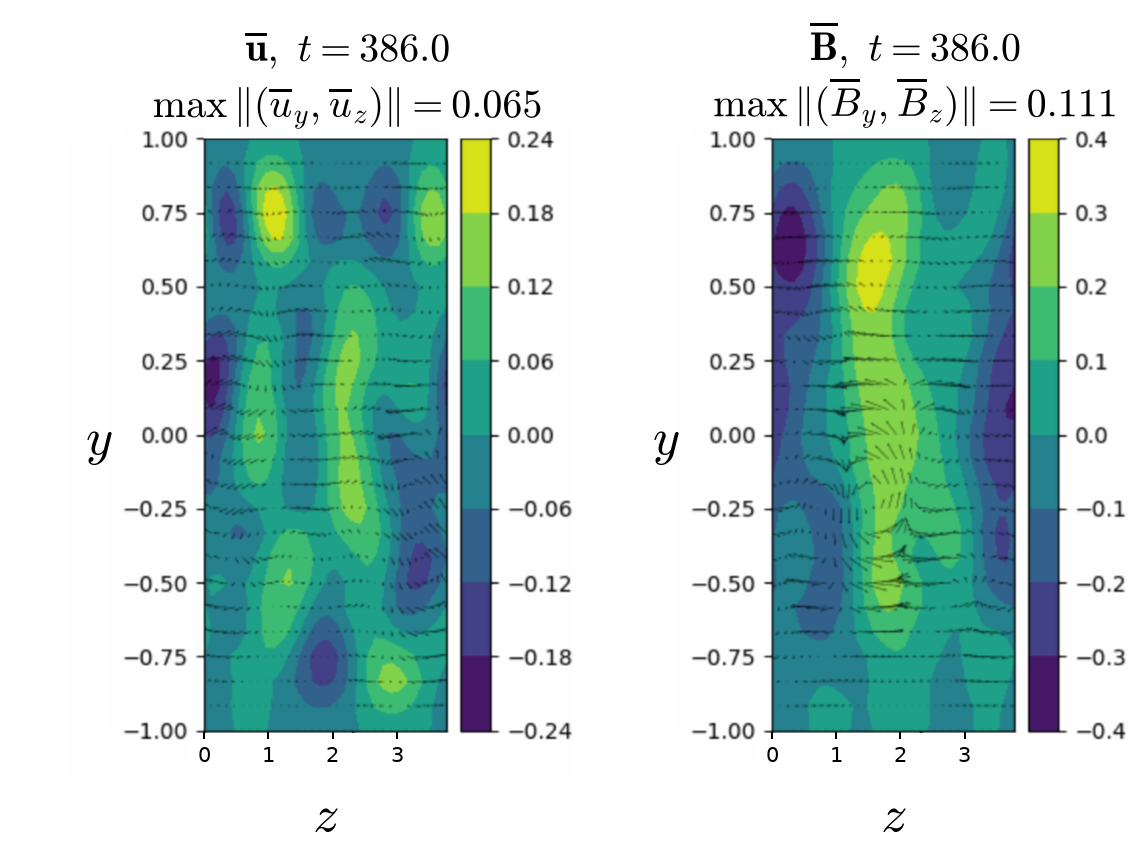}
            \label{subfig:a}%
        }\hfill
        \subfloat[]{%
            \includegraphics[width=.48\linewidth]{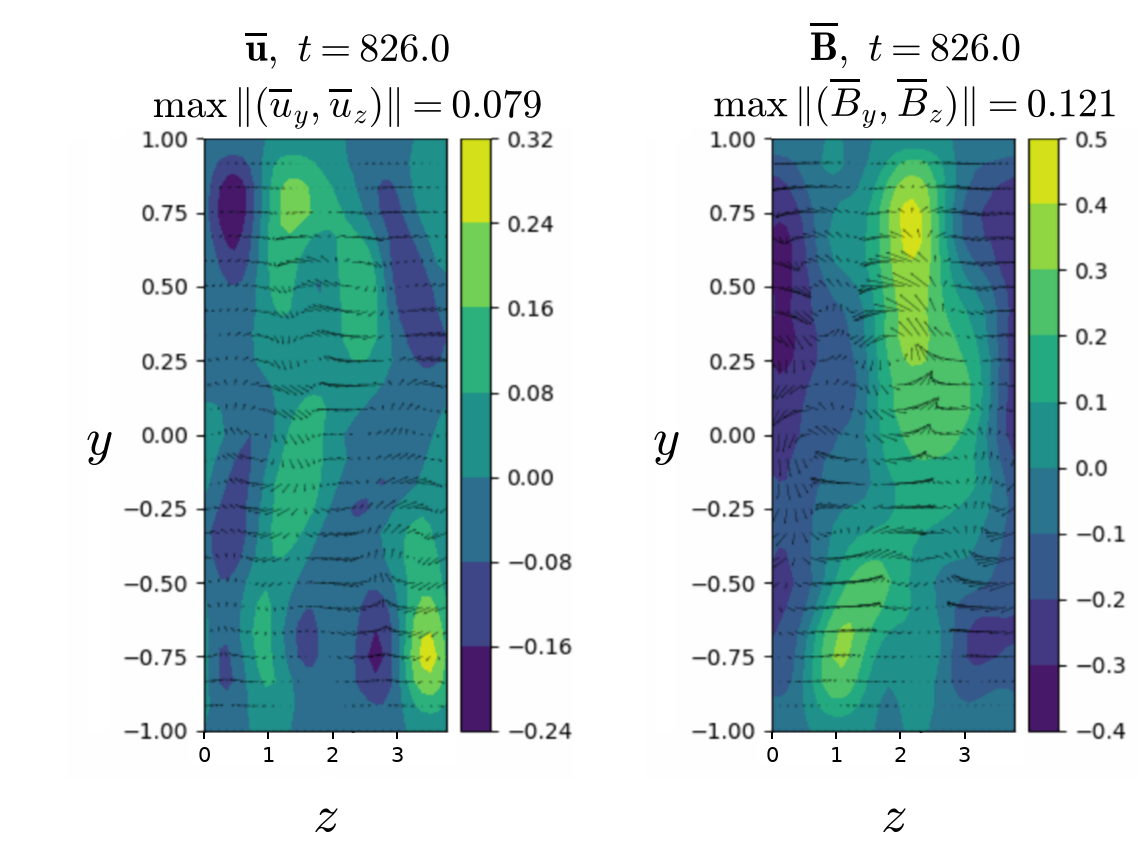}
            \label{subfig:b}%
        }\\
        \subfloat[]{%
            \includegraphics[width=.48\linewidth]{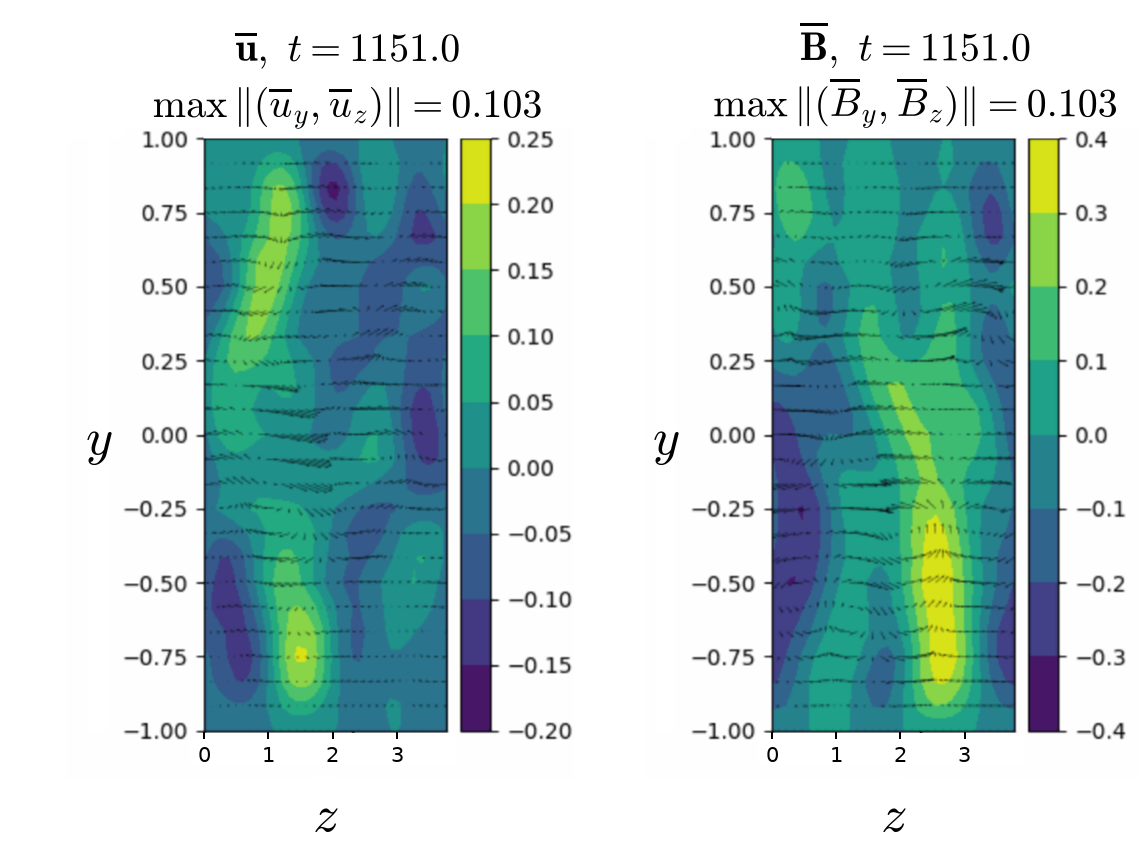}
            \label{subfig:c}%
        }\hfill
        \subfloat[]{%
            \includegraphics[width=.48\linewidth]{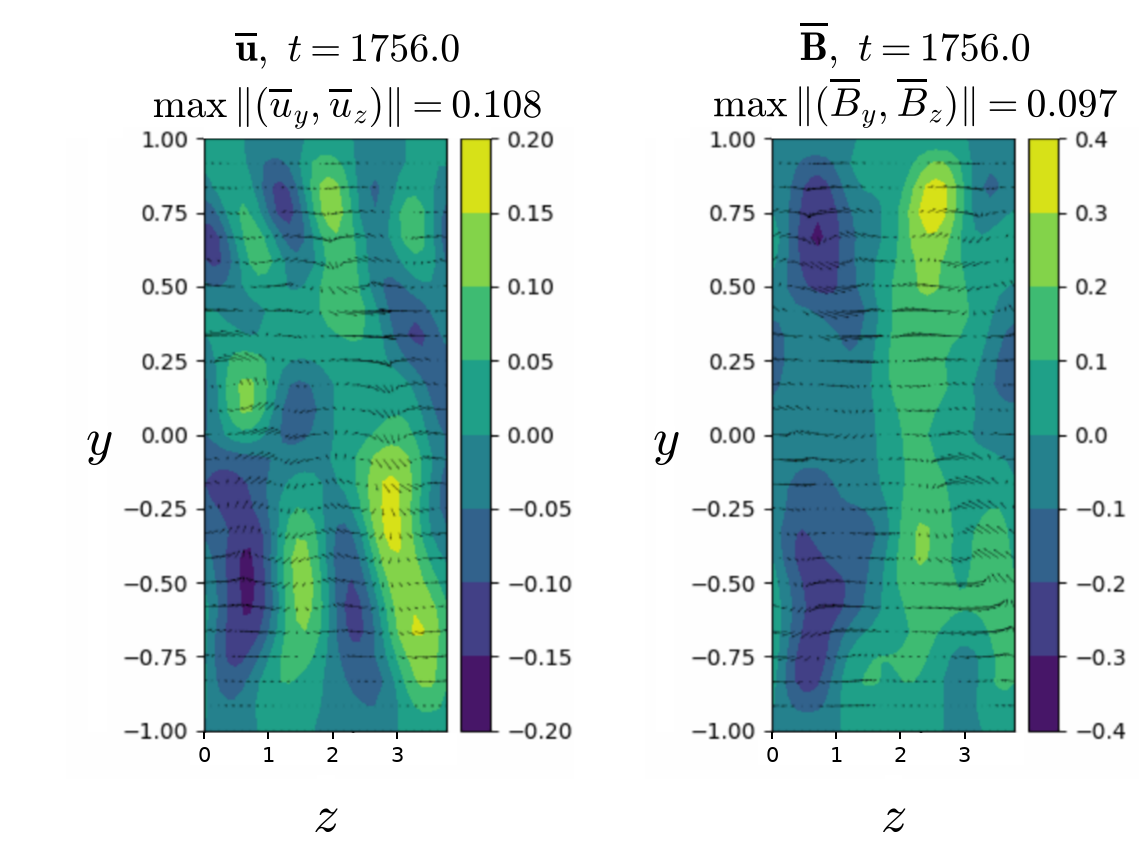}
            \label{subfig:d}%
        }
        
\caption{Snapshots of self sustaining turbulent equilibria maintaining finite amplitude components in both  mean velocity $\overline{\bv{u}}$ and mean magnetic field $\overline{\bv{B}}$. For each snapshot, on the left: streak velocity $\overline{u}_{xs}=\overline{u}_x-[\overline{u}_x]_z$ (contours) and roll velocity $(\overline{u}_{y},\overline{u}_z)$ (vectors). On the right: streak magnetic field $\overline{B}_{xs}=\overline{B}_x-[\overline{B}_x]_z$ (contours) and roll  magnetic field $(\overline{B}_y,\overline{B}_z)$ (vectors). (Note the approximate constancy with time in the amplitude of the magnetic field component.) Parameters: $Re=400$, $Re_m=600$ ($Pr_m=1.5>Pr_{mc}\approx1.25$; cf.\ figure~\ref{fig:Prmbifur}), $\epsilon=0$.}
\label{fig:uBSSP}
\end{figure}

Consistent with this result, at $Re=400$, $Pr_m=1.5$ (marked by a square in figure \ref{fig:Prmbifur}), a self sustaining  turbulent RSS is maintained that supports both a coherent mean velocity and magnetic field component. Snapshots of this self sustaining state are shown in figure \ref{fig:uBSSP}.

It is also important to point out that while we obtained $Pr_{mc} \approx1.25>1$ for $Re=400$, $Pr_{mc}$ is a function of Reynolds number and $Pr_{mc}$ is not necessarily greater than unity. For example, at $Re=600$, $Pr_{mc}\approx 0.83<1$, which shows that self sustaining dynamos can exist in our system at both $Pr_m>1$ and $Pr_m<1$.

\section{Mechanisms maintaining the velocity component in turbulent MHD RSS}\label{velmaintsec}
We now consider the physical mechanisms maintaining the RSS in turbulent MHD Couette flow and relate these
to the mechanisms maintaining the RSS in wall-bounded shear flows. 
Unlike the case of the RSS in wall bounded shear flows, 
in MHD Couette turbulence the RSS is a composite structure consisting of interacting components in both the velocity $\overline{\bv{u}}$ and the magnetic field $\overline{\bv{B}}$.
For each of the maintenance diagnostics that follow we show results from the SSD together with results from direct numerical simulation (DNS) of the full MHD equations \eqref{nondimNS}--\eqref{nondimind} for comparison. The DNS is performed in the same domain, $1.75\pi\times2\times1.2\pi$, with the same boundary conditions and at $Re=Re_m=400$, with a resolution of $32\times33\times32$. The diagnostics are shown as time series of the terms averaged over $y$ and $z$.

We begin by extending previous studies on velocity field RSS maintenance in wall bounded shear flow turbulence \citep{Farrell 2016} to the MHD case.  In that study the mechanisms underlying RSS dynamics and equilibration were examined using
balance equations for the RSS streak velocity, $\overline{u}_{xs}\equiv\overline{u}_x-[\overline{u}_x]_z$, and roll vorticity, $\overline{\omega}_{xr}\equiv\overline{\omega}_x-[\overline{\omega}_x]_z$,  where the streamwise component of vorticity is defined as $\overline{\omega}_x=-\partial_z \overline{u}_y+\partial_y \overline{u}_z$.\\

We first study maintenance of the streak velocity component of the RSS, $\overline{u}_{xs}$, for which the balance equation is:\\
\begin{equation}
\begin{split}
\partial_t \overline{u}_{xs}=-(\overline{u}_y(\partial_y \overline{u}_x)-[\overline{u}_y(\partial_y \overline{u}_x)]_z)-(\overline{u}_z(\partial_z \overline{u}_x)-[\overline{u}_z(\partial_z \overline{u}_x)]_z)\\-([u_y'(\partial_y u_x')]_{x}-[u_y'(\partial_y u_x')]_{x,z})-([u_z'(\partial_z u_x')]_{x}-[u_z'(\partial_z u_x')]_{x,z})\\+ (\overline{B}_y(\partial_y\overline{B}_x)-[\overline{B}_y(\partial_y\overline{B}_x)]_{z})+(\overline{B}_z(\partial_z\overline{B}_x)-[\overline{B}_z(\partial_z\overline{B}_x)]_{z})
\\
+([B'_y(\partial_yB'_x)]_{x}-[B'_y(\partial_yB'_x)]_{x,z})+([B'_z(\partial_zB'_x)]_{x}-[B'_z(\partial_zB'_x)]_{x,z})
\\+\Delta_1 
\frac{\overline{u}_{xs}}{Re}.
\end{split}
\end{equation}\\
in which $\Delta_1\equiv\partial_{yy}+\partial_{zz}$ is the Laplacian in the cross-stream--spanwise plane, which acts on the streamwise-mean fields.

Because streaks of both signs occur,  obtaining a linear measure of streak
forcing requires multiplying each term by the local $sign(\overline{u}_{xs})$. The resulting sign-weighted diagnostics of the physically distinct terms in the dynamics maintaining $\overline{u}_{xs}$, $I_A$--$I_E$, are defined in appendix~\ref{app:Iterms}.

The components of streak velocity forcing are identified as lift-up, $(I_A)$, Reynolds stress divergence, $(I_B)$,
magnetic tension force due to mean magnetic field, $(I_C)$, divergence of Maxwell stress, $(I_D)$,
and viscous damping, ($I_E$). 

\begin{figure}
\centering{
\begin{subfigure}{0.8\textwidth} \caption{}
\includegraphics[width=\linewidth]{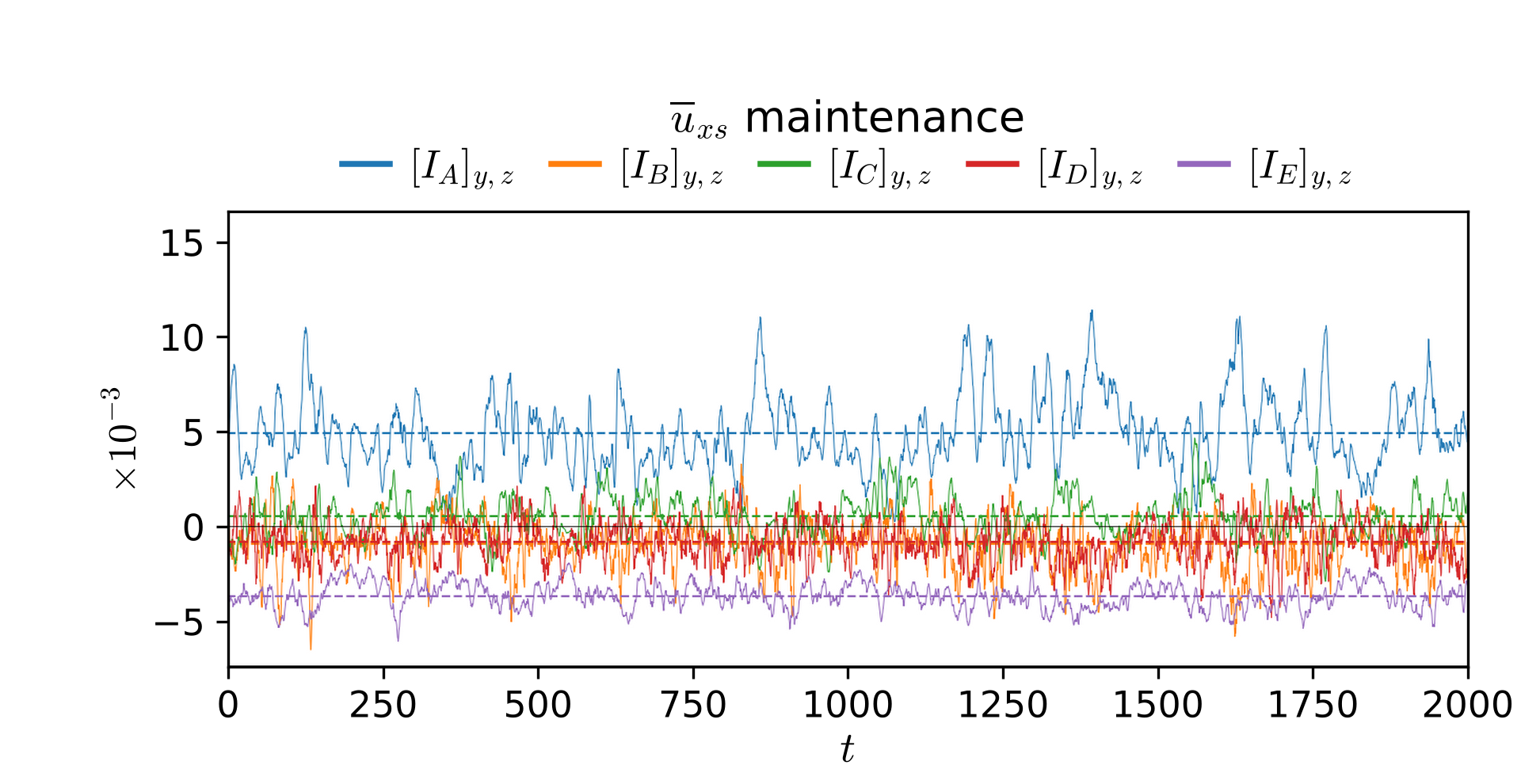}
\end{subfigure}
\begin{subfigure}{0.8\textwidth} \caption{}
\includegraphics[width=\linewidth]{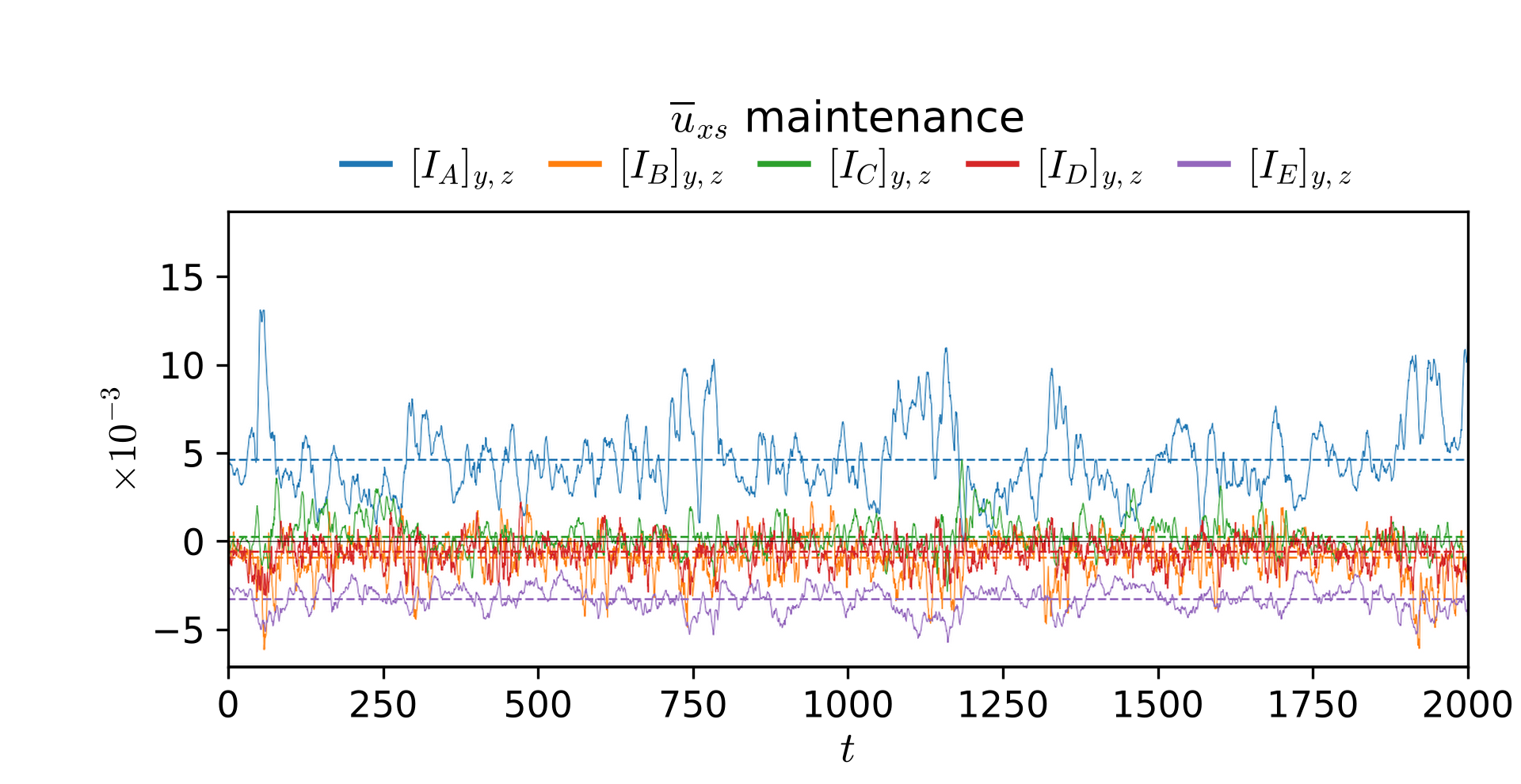}
\end{subfigure}
}
\caption{Equilibrium balance for the velocity streak, $\overline{u}_{xs}$, maintenance showing the lift-up term, $I_A(y, z, t)$, Reynolds stress divergence term, $I_B(y, z, t)$,  magnetic tension force due to mean magnetic field $I_C(y,z,t)$,  divergence of Maxwell stress $I_D(y,z,t)$, and diffusion $I_{E}(y,z,t)$, averaged in the spanwise direction, $z$, and the wall normal direction, $y$. (a) SSD and (b) DNS, both at $Re=Re_m=400$; in the SSD $\epsilon=3.65$.}
\label{Usturbulence}
\end{figure}

As an example of the use of balance diagnostics, we show the balance for the case of turbulence with parameters $Re=400$, $Re_m=400$, $\epsilon=3.65$ shown in figure \ref{fig:turbulentsnapshot}.  Time series of the spatially averaged diagnostics, $I_A,I_B,I_C,I_D,I_E$,  shown in figure \ref{Usturbulence}, reveal the dominant balance is between the lift-up process, $I_A$, and viscous dissipation, $I_E$. Reynolds stress divergence, $I_B$, together with the Maxwell stress divergence, $I_D$, are relatively small quantitatively; however, these second cumulant forcings are crucial components of the mechanism maintaining the statistical equilibrium through inducing feedback regulation on the first cumulant.
 Similar to the streak velocity force balance in wall bounded shear flow, the balance is between a positive contribution from lift-up, $I_A$, opposed by negative contributions from dissipation, $I_E$,  and fluctuation-fluctuation stresses. The difference is that down gradient fluctuation-fluctuation stress divergence comes from not only Reynolds stress divergence, $I_B$,  but also divergence of Maxwell stress, $I_D$. The negative contributions of $I_B$ and $I_D$ indicate that both the fluctuation velocity field, $\bv{u'}$,  and  the fluctuation magnetic field, $\bv{B'}$, extract energy from the streak  velocity, $\overline{u}_{xs}$. This occurs because the streak velocity shears the background turbulence, in both the velocity and the magnetic field, so that the fluctuation velocity and magnetic field become systematically correlated in such a way that their Reynolds and Maxwell stresses act down-gradient on the streak and extract energy from it. Results from the SSD (figure \ref{Usturbulence}a) and the DNS (figure \ref{Usturbulence}b) agree.  
 
 We now consider the roll vorticity, $\overline{\omega}_{xr}$, maintenance using as a diagnostic the roll vorticity forcing.
The roll vorticity forcing balance is:

\begin{equation}
\begin{split}
\partial_t \overline{\omega}_{xr}=-((\overline{u}_y\partial_y + \overline{u}_z \partial_z)\overline{\omega}_{x}-[(\overline{u}_y\partial_y + \overline{u}_z \partial_z) \overline{\omega}_{x}]_z) +(\partial_{zz}-\partial_{yy})([u_y'u_z']_x-[u_y'u_z']_{x,z})\\- \partial_{yz}([u_z'u_z']_x-[u_y'u_y']_x)+(\partial_{yy}-\partial_{zz})(\overline{B}_y\overline{B}_z-[\overline{B}_y\overline{B}_z]_z)+\partial_{yz}(\overline{B}_z\overline{B}_z-\overline{B}_y\overline{B}_y)\\+   (\partial_{yy}-\partial_{zz})([B'_yB'_z]_x-[B'_yB'_z]_{x,z})+\partial_{yz}([B'_zB'_z]_x-[B'_yB'_y]_x)+\Delta_1\frac{\overline{\omega}_{xr}}{Re}.
\end{split}
\end{equation}\\
As in the above case for the streak velocity, the sign-weighted diagnostics of the physically distinct terms in the dynamics maintaining $\overline{\omega}_{xr}$, $I_F$--$I_J$, are defined in appendix~\ref{app:Iterms}.

\begin{figure}
\centering{
\begin{subfigure}{0.8\textwidth} \caption{}
\includegraphics[width=\linewidth]{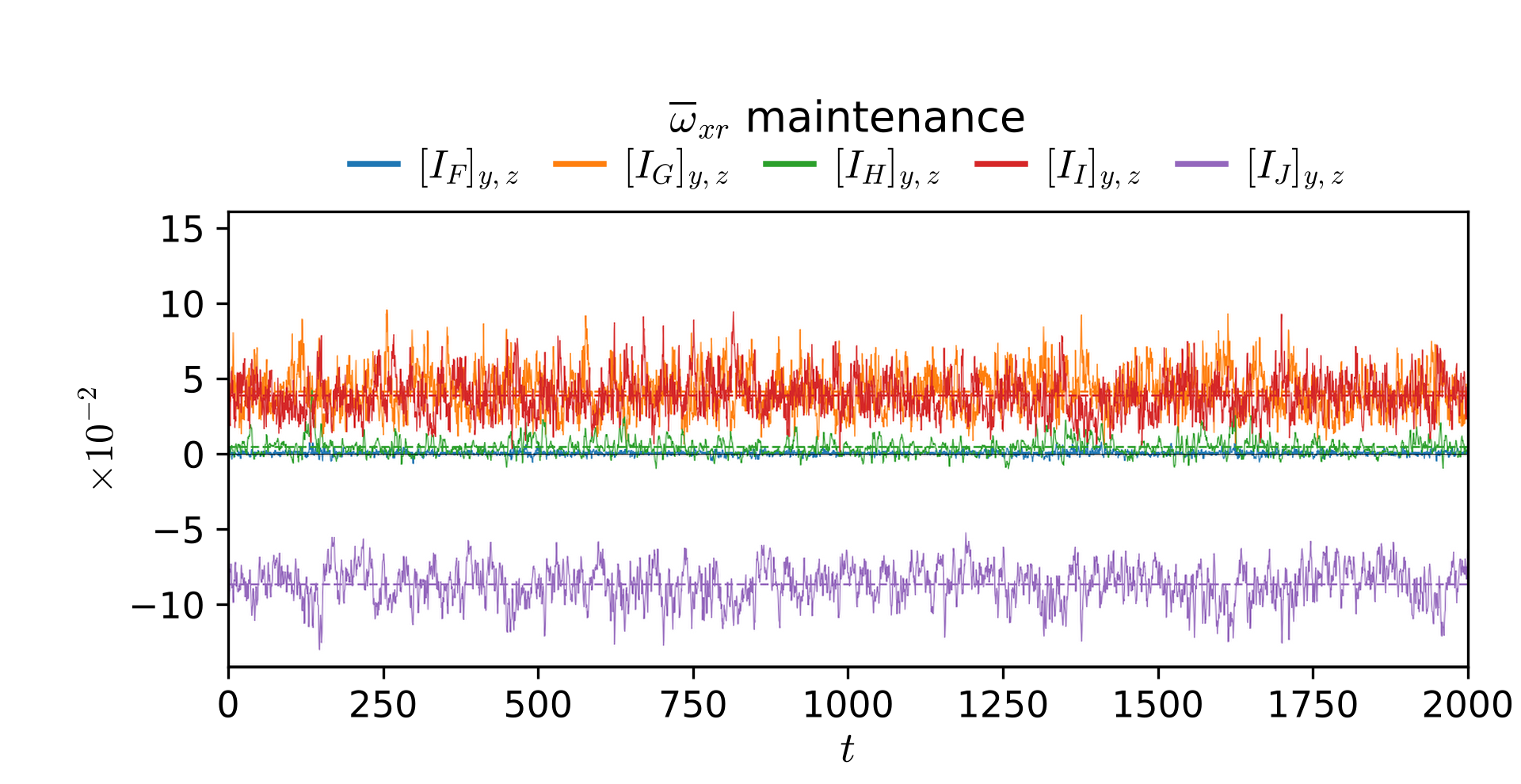}
\end{subfigure}
\begin{subfigure}{0.8\textwidth} \caption{}
\includegraphics[width=\linewidth]{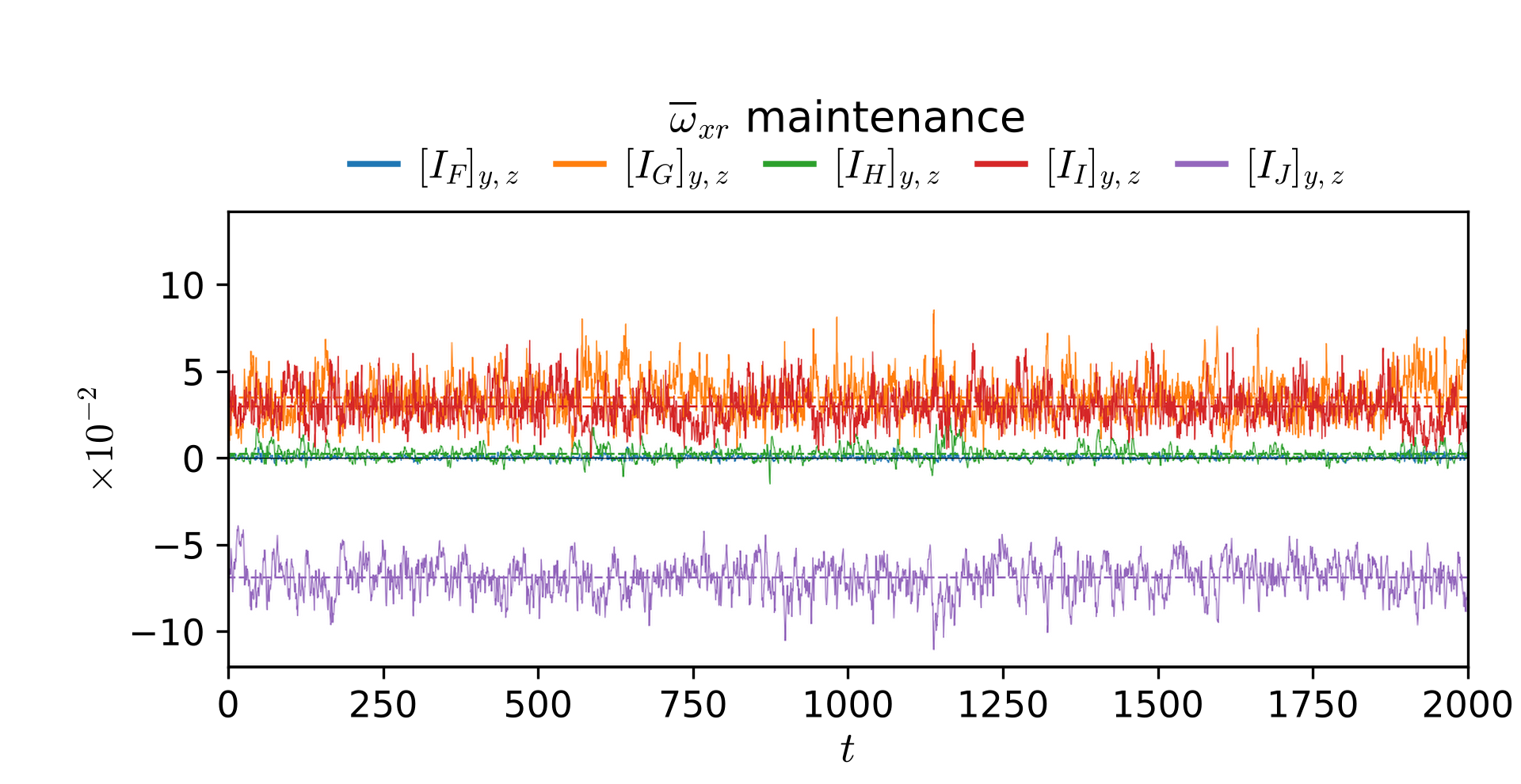}
\end{subfigure}
}
\caption{Equilibrium balance for streamwise vorticity (streamwise current density),  $\overline{\omega}_{xr}$,  containing mean advection term, $I_F(y, z, t)$, vorticity induced by Reynolds stress, $I_G(y, z, t)$, magnetic tension vorticity induced by mean magnetic field,  $I_H(y,z,t)$, and vorticity induced by divergence of Maxwell stress, $I_I(y,z,t)$, and diffusion $I_{J}(y,z,t)$, averaged in the spanwise direction, $z$, and the wall normal direction, $y$. (a) SSD and (b) DNS, both at $Re=Re_m=400$; in the SSD $\epsilon=3.65$.}
\label{Omegarturbulence}
\end{figure}
The components of roll vorticity forcing are identified as advection by mean velocity, ($I_F$), vorticity induced by Reynolds stress, ($I_G$), vorticity induced by magnetic tension force of mean magnetic field, $(I_H)$, vorticity induced by Maxwell stress, $(I_I)$, and dissipation, $(I_J)$.
Time series of the spatially averaged diagnostics of $I_F, I_G , I_H , I_I , I_J$ shown in figure \ref{Omegarturbulence} reveal that the dominant balance maintaining the roll vorticity is between the positive contribution from Reynolds stress divergence, $I_G$ and Maxwell stress divergence, $I_I$, and the negative contribution from dissipation, $I_J$.  Similar to the roll vorticity force balance in wall bounded shear
flow, the dominant balance is between the positive contributions from fluctuation-fluctuation terms  opposed
by a negative contribution from dissipation, $I_J$. The
difference is that the positive contribution from the fluctuation-fluctuation terms come from
not only the Reynolds stress term, $I_G$, but also the Maxwell stress term, $I_I$. Physically, the streak velocity shears the background turbulence, in both the velocity and the magnetic field, which makes the fluctuation velocity and magnetic field systematically correlated. The Reynolds stresses and Maxwell stresses of these correlated fluctuations drive the streamwise vorticity against dissipation. The dominant balances obtained from the SSD (figure \ref{Omegarturbulence}a) and the DNS (figure \ref{Omegarturbulence}b) agree. 

These results show that the mechanisms maintaining $\overline{u}_{xs}$ and $\overline{\omega}_{xr}$ in MHD turbulence are similar to those maintaining these components in wall-bounded shear flow, the main difference being the contribution by Maxwell stress divergence contributing a negative forcing by $I_D$ to $\overline{u}_{xs}$, indicative of transfer of streak energy to the fluctuation magnetic field, and a positive contribution by $I_I$ to $\overline{\omega}_{xr}$, indicative of forcing of roll circulation by Maxwell stress.\\

\section{Mechanisms maintaining the magnetic field in turbulent MHD RSS}\label{magmaintsec}
We turn now to examine in a similar manner the physical mechanisms maintaining the magnetic field component of the RSS.
The balance equation for maintenance of the streak magnetic field, $\overline{B}_{xs}=\overline{B}_x-[\overline{B}_x]_z$, is:\\

\begin{equation}
\begin{aligned}
\begin{split}
\partial_t \overline{B}_{xs}=((\partial_y \overline{u}_x)\overline{B}_y-[(\partial_y \overline{u}_x)\overline{B}_y]_{z})+((\partial_z\overline{u}_x)\overline{B}_z-[(\partial_z\overline{u}_x)\overline{B}_z]_z) \\+([(\partial_x u'_x)B'_x]_x-[(\partial_x u'_x)B'_x]_{x,z})+([(\partial_y u'_x)B'_y]_x-[(\partial_y u'_x)B'_y]_{x,z})\\+([(\partial_z u'_x)B'_z]_x-[(\partial_z u'_x)B'_z]_{x,z})-((\partial_y \overline{B}_x)\overline{u}_y-[(\partial_y \overline{B}_x)\overline{u}_y]_{z})\\-((\partial_z \overline{B}_x)\overline{u}_z-[(\partial_z \overline{B}_x)\overline{u}_z]_z)-([(\partial_x B'_x)u'_x]_x-[(\partial_x B'_x)u'_x]_{x,z})\\-([(\partial_y B'_x)u'_y]_x-[{(\partial_y B'_x)u'_y}]_{x,z})-([(\partial_z B'_x)u'_z]_x-[{(\partial_z B'_x)u'_z}]_{x,z})\\+\frac{1}{Re_m}\Delta_1 \overline{B}_{xs}
\end{split}
\end{aligned}
\end{equation}\\

As in the above case for the streak velocity, the sign-weighted diagnostics of the physically distinct terms in the dynamics maintaining the streak magnetic field $\overline{B}_{xs}$, $I_{BA}$--$I_{BE}$, are defined in appendix~\ref{app:Iterms}.

\begin{figure}
\centering{
\begin{subfigure}{0.8\textwidth} \caption{}
\includegraphics[width=\linewidth]{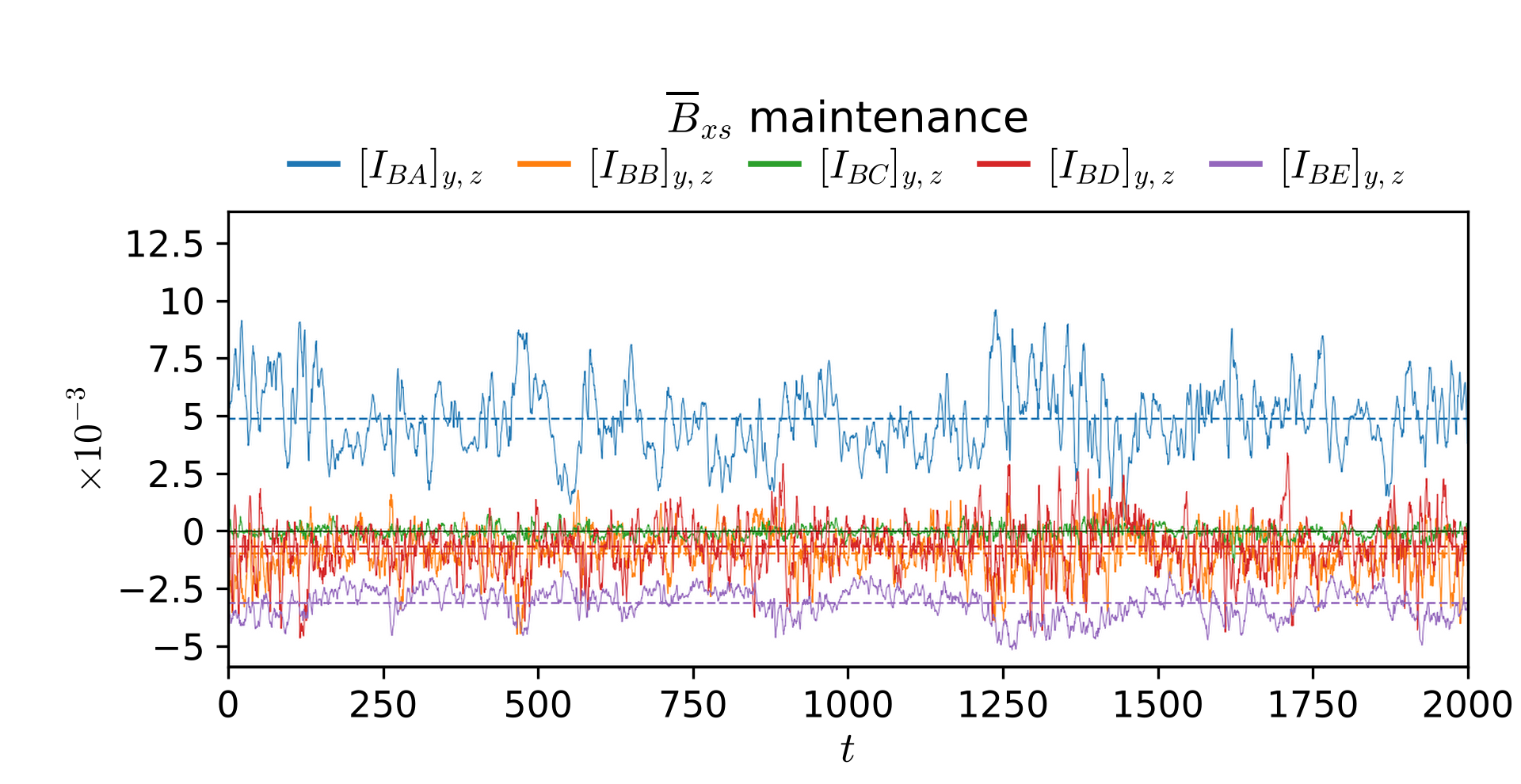}
\end{subfigure}
\begin{subfigure}{0.8\textwidth} \caption{}
\includegraphics[width=\linewidth]{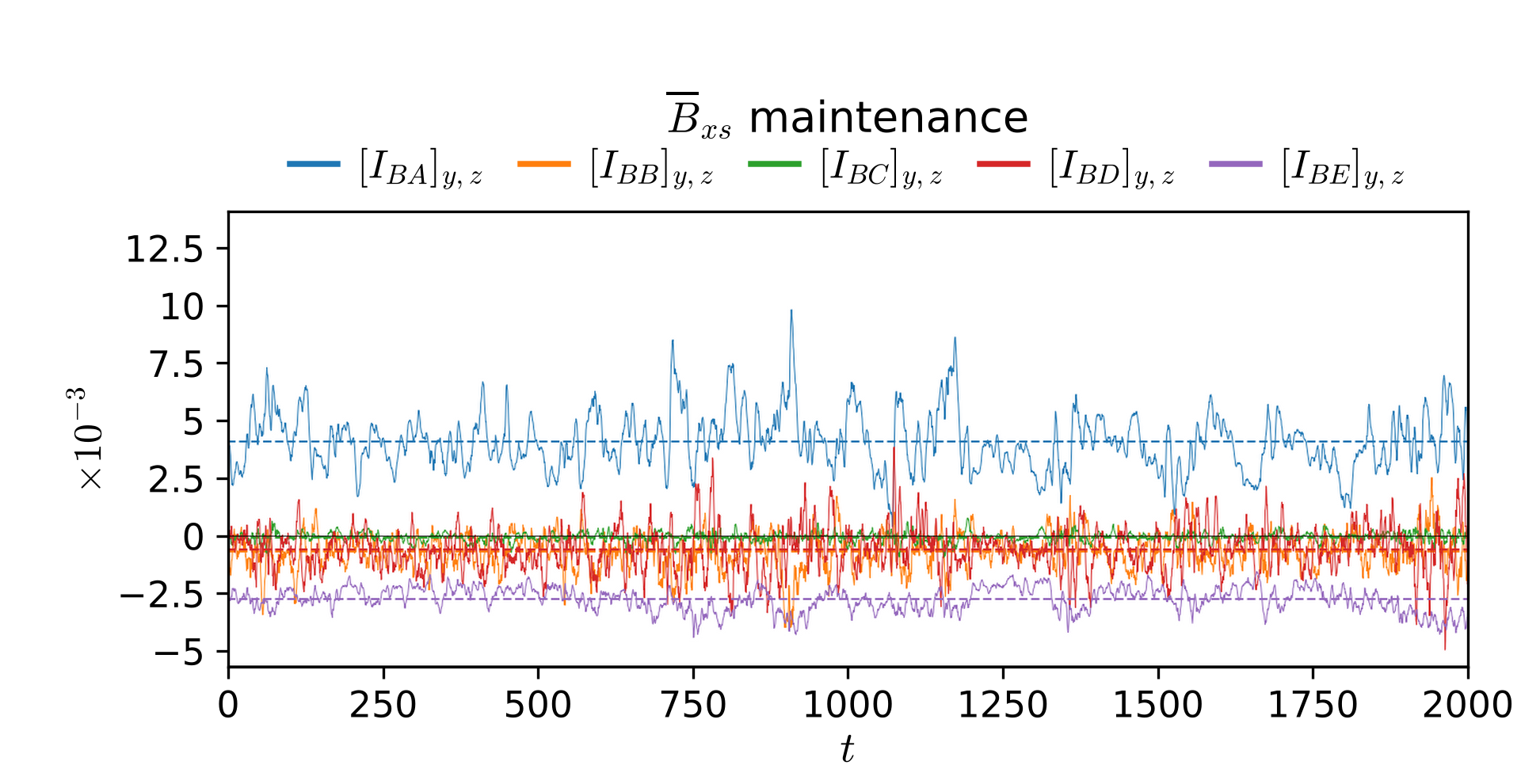}
\end{subfigure}
}
\caption{Equilibrium balance of magnetic field streak, $B_{xs}$, in  RSS turbulence. Shown are mean field tilting, $I_{BA}$, fluctuation-fluctuation tilting, $I_{BB}$, mean field advection, $I_{BC}$, fluctuation-fluctuation advection, $I_{BD}$, and diffusion $I_{BE}$, averaged in the spanwise direction, $z$, and the wall normal direction, $y$. (a) SSD and (b) DNS, both at $Re=Re_m=400$; in the SSD $\epsilon=3.65$.}
\label{Bsmaintenance}
\end{figure}

The components of  streak magnetic field forcing are identified as mean tilting, $(I_{BA})$, fluctuation tilting, $(I_{BB})$,
mean advection, $(I_{BC})$, fluctuation advection, $(I_{BD})$,
and diffusive damping, ($I_{BE}$). Time series of the spatially averaged diagnostics of $I_{BA}$, $I_{BB}$, $I_{BC}$, $I_{BD}$, and $I_{BE}$  shown for our case of turbulence with $Re=400, Re_m=400$, $   \epsilon=3.65$ in figure \ref{Bsmaintenance} reveal that the dominant balance is between mean tilting, $I_{BA}$, and dissipation, $I_{BE}$. While fluctuation tilting, $I_{BB}$, and fluctuation advection, $I_{BD}$, are relatively small quantitatively, these terms are important in regulating the magnetic field at its statistical equilibrium. While streak velocity is maintained against dissipation primarily by lift up, $I_{A}$, streak magnetic field is maintained primarily by mean tilting, $I_{BA}$. The negative contributions of fluctuation tilting, $I_{BB}$, and fluctuation advection, $I_{BD}$, indicate that the correlation between the fluctuation velocity field, $\bv{u'}$,  and  the fluctuation magnetic field, $\bv{B'}$, results in systematic extraction of energy from the streak  magnetic field, $\overline{B}_{xs}$, which energy is transferred to both $\bv{u'}$ and $\bv{B'}$. This occurs because the streak magnetic field shears the background turbulence, so that the fluctuation velocity and magnetic field become systematically correlated in such a way that their electromotive force acts down-gradient on the streak magnetic field and extracts energy from it. This transfer is one component of the fluctuation feedback, arising from the electromotive force produced by the RSS-organized correlation of the velocity and magnetic field fluctuations. The dominant balances obtained from the SSD (figure \ref{Bsmaintenance}a) and the DNS (figure \ref{Bsmaintenance}b) agree.

 We now examine maintenance of the roll magnetic field by first forming the curl of the magnetic field in the y-z plane, $\overline{\zeta_x}\equiv-\partial_z\overline{B_y}+\partial_y \overline{B_z}$, and take as our diagnostic for the roll magnetic field the quantity $\overline{\zeta_x}$ after removal of its spanwise average, 
$\overline{\zeta}_{xr}\equiv\overline{\zeta}_{x}-[\overline{\zeta}_{x}]_z$. 
The forcing of this quantity, which can be identified as the current density flowing into the y-z plane, has balance:

\begin{equation}
\begin{aligned}
\begin{split}
\partial_t\overline{\zeta}_{xr}=\partial_y((\partial_y\overline{u}_z)\overline{B}_y-[(\partial_y\overline{u}_z)\overline{B}_y]_z)+\partial_y((\partial_z\overline{u}_z)\overline{B}_z-[(\partial_z \overline{u}_z) \overline{B}_z]_{z})-\partial_{z}((\partial_y \overline{u}_y) \overline{B}_y)\\-\partial_z((\partial_z \overline{u}_y) \overline{B}_z)
+\partial_y([ (\partial_x u'_z)B'_x]_x-[(\partial_x u'_z)B'_x]_{x,z})+\partial_y([ (\partial_y u'_z)B'_y]_x-[ (\partial_y u'_z)B'_y]_{x,z})\\+\partial_y([ (\partial_z u'_z )B'_z]_x-[ (\partial_z u'_z )B'_z]_{x,z})-\partial_{z}([ (\partial_x u'_y) B'_x]_x)-\partial_z([ (\partial_y u'_y) B'_y]_x)\\-\partial_z(  [(\partial_z u'_y) B'_z]_x)
+\partial_z((\partial_y \overline{B}_y)\overline{u}_y)+\partial_z((\partial_z\overline{B}_y)\overline{u}_z)-\partial_y( (\partial_y\overline{B}_z)\overline{u}_y-[(\partial_y\overline{B}_z)\overline{u}_y]_z)\\-\partial_y((\partial_z\overline{B}_z)\overline{u}_z-[(\partial_z\overline{B}_z)\overline{u}_z]_{z})+\partial_z([ (\partial_x B'_y)u'_x]_{x})+\partial_z([ (\partial_y B'_y)u'_y]_{x})\\+\partial_z([(\partial_z B'_y)u'_z]_x)-\partial_y([(\partial_x B'_z)u'_x]_x-[(\partial_x B'_z)u'_x]_{x,z})\\-\partial_y([(\partial_y B'_z)u'_y]_x-[(\partial_y B'_z)u'_y]_{x,z})-\partial_y([(\partial_z B'_z)u'_z]_x-[(\partial_z B'_z)u'_z]_{x,z})+\frac{1}{Re_m}\Delta_{1}\overline{\zeta}_{xr}
\end{split}
\end{aligned}
\end{equation}

As in the above case for the streak magnetic field, the sign-weighted diagnostics of the physically distinct terms in the dynamics maintaining $\overline{\zeta}_{xr}$, $I_{BF}$--$I_{BJ}$, are defined in appendix~\ref{app:Iterms}.

These components of roll current density forcing are identified as the current density in the $x$ direction induced by mean tilting and stretching, ($I_{BF}$), fluctuation tilting and stretching, ($I_{BG}$),  mean advection, $(I_{BH})$, fluctuation advection, $(I_{BI})$, and dissipation, $(I_{BJ})$.

\begin{figure}
\centering{
\begin{subfigure}{0.8\textwidth} \caption{}
\includegraphics[width=\linewidth]{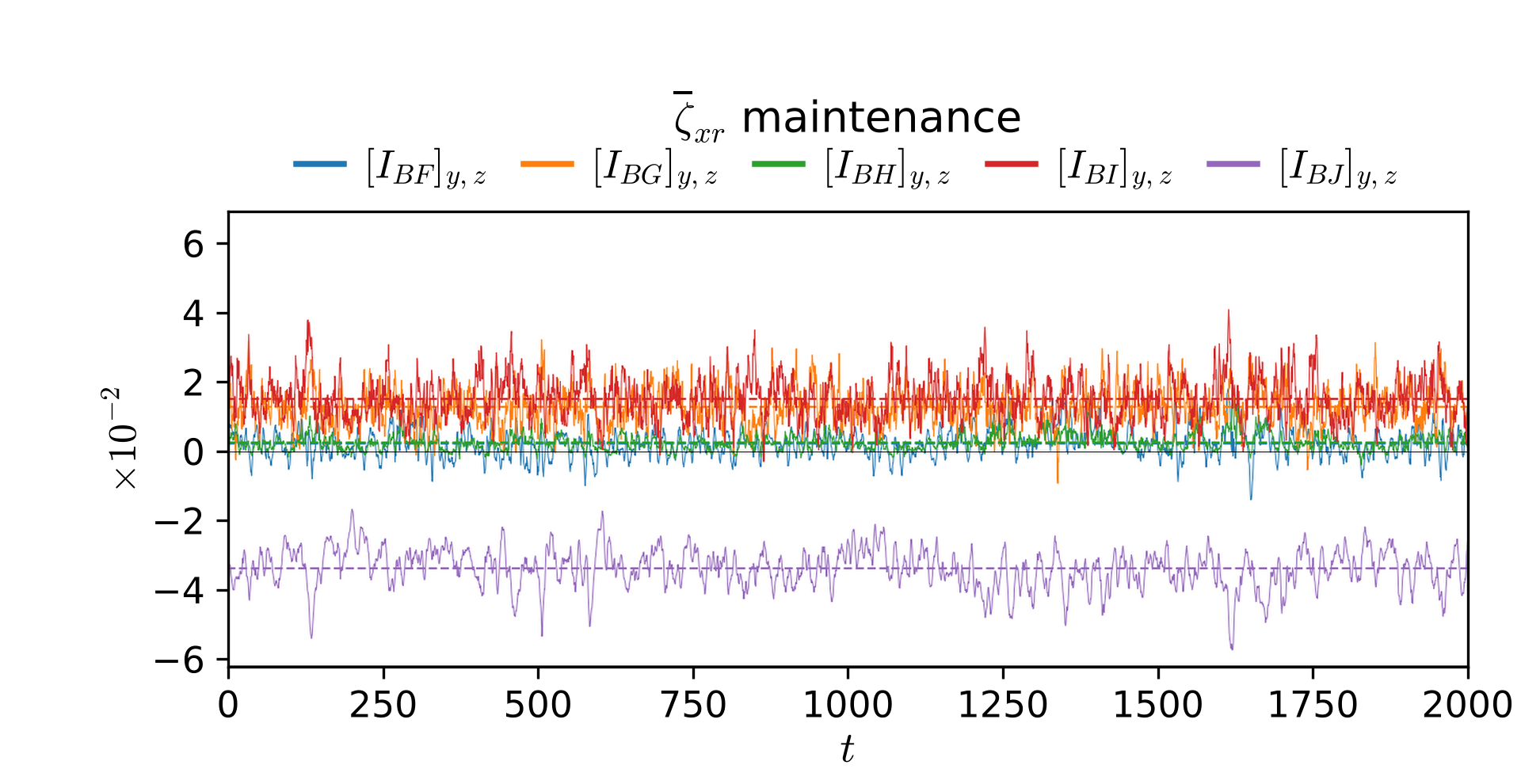}
\end{subfigure}
\begin{subfigure}{0.8\textwidth} \caption{}
\includegraphics[width=\linewidth]{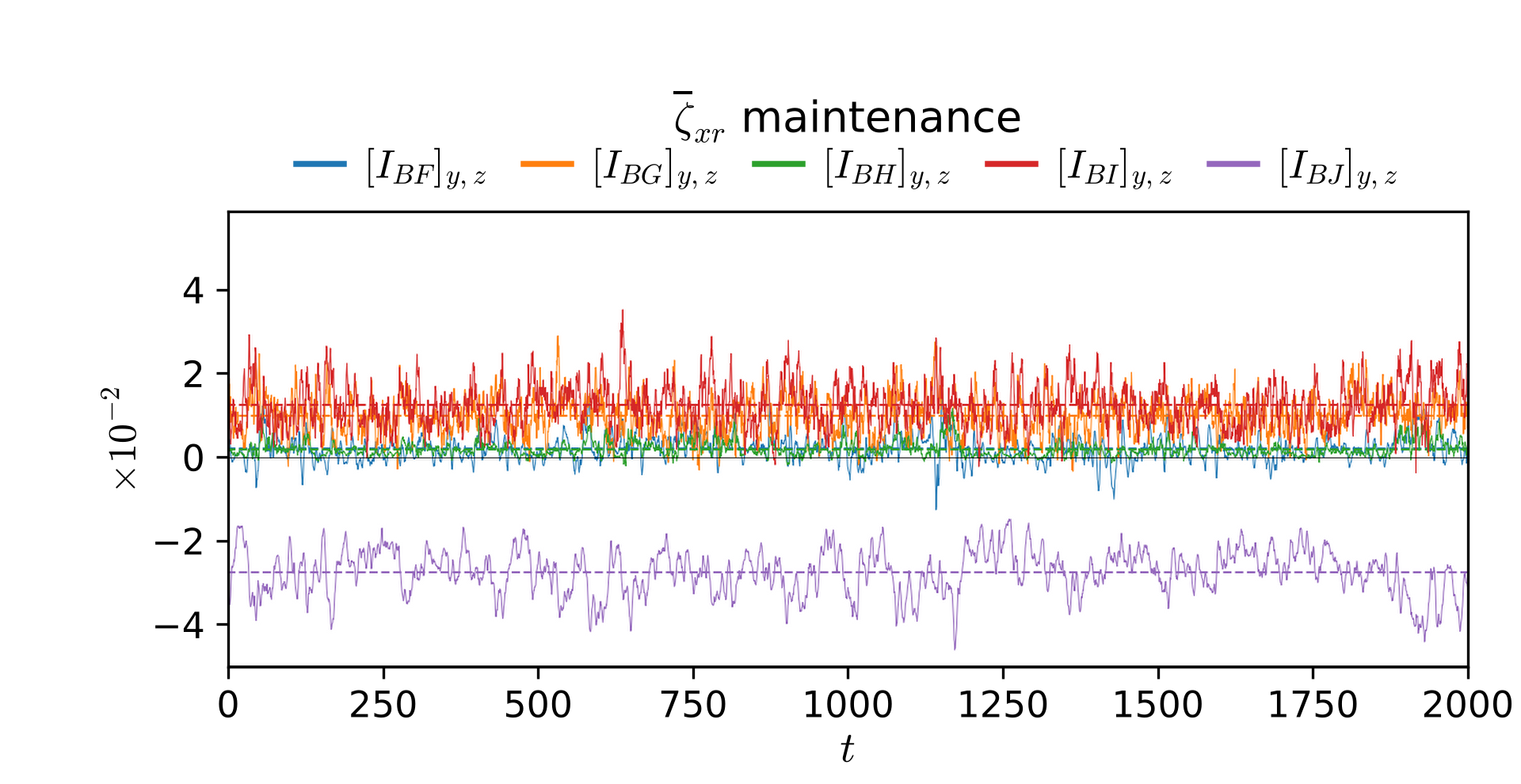}
\end{subfigure}
}
\caption{
Equilibrium balance of magnetic field roll vorticity density in  RSS turbulence.  Shown are the mean tilting term, $I_{BF}$, fluctuation-fluctuation tilting, $I_{BG}$, mean advection of magnetic vorticity, $I_{BH}$, fluctuation-fluctuation advection, $I_{BI}$, and diffusion of magnetic field vorticity, $I_{BJ}$, averaged in the spanwise direction, $z$, and the wall normal direction, $y$. (a) SSD and (b) DNS, both at $Re=Re_m=400$; in the SSD $\epsilon=3.65$.
}
\label{OmegaBrmaintenance}
\end{figure}

Time series of the spatially averaged diagnostics of $I_{BF}$, $I_{BG}$, $I_{BH}$, $I_{BI}$, and $I_{BJ}$ are shown in figure \ref{OmegaBrmaintenance}. 
These diagnostics reveal that the dominant balance maintaining the roll current density is between the positive contribution from fluctuation tilting and stretching, $I_{BG}$, and fluctuation advection, $I_{BI}$, and the negative contribution from dissipation, $I_{BJ}$. Physically, the cross correlations between the fluctuation velocity and magnetic field are systematically organized so that their electromotive force drives the current density in the $x$ direction, maintaining the roll magnetic field against dissipation. Results from the SSD (figure \ref{OmegaBrmaintenance}a) and the DNS (figure \ref{OmegaBrmaintenance}b) agree. 

Comparing this roll current density  force balance with the force balance of the roll vorticity we note that in both cases the dominant balance is between the positive contributions from fluctuation-fluctuation terms  opposed
by a negative contribution from dissipation. 
Despite differences in physical interpretations of individual terms in these balances,
our results show that the mechanisms maintaining the magnetic field components, $\overline{B}_{xs}$ and $\overline{\zeta}_{xr}$, in turbulent MHD RSS are remarkably parallel to the mechanisms maintaining the velocity components, $\overline{u}_{xs}$ and $\overline{\omega}_{xr}$.

Summarizing, the streak velocity is produced by lift up and the streak magnetic field is produced by tilting. Maintenance of both lift up and tilting require maintenance of the respective curl components $\overline{\omega}_{xr}$ and $\overline{\zeta}_{xr}$. These are both maintained against dissipation by positive contributions produced by fluctuation-fluctuation terms. We find two distinct fluctuation-fluctuation terms exist for each, $I_{G}$ and $I_{I}$ for $\overline{\omega}_{xr}$ maintenance and $I_{BG}$ and $I_{BI}$ for $\overline{\zeta}_{xr}$ maintenance. 

The focus of the preceding and this section was on understanding the maintenance of the components of the RSS contained in the mean velocity field $\overline{\bv{u}}$ and mean magnetic field $\overline{\bv{B}}$.  We next turn our attention to the dynamics of the fluctuation components contained in the  covariance.

\section{Maintenance of the fluctuation turbulent kinetic and magnetic energy components}

Fluctuation turbulent kinetic energy (TKE) and turbulent magnetic energy (TME) are defined as:\\

\begin{equation}
TKE(t)=\frac{1}{2}[u'^2_x+u'^2_y+u'^2_z]_{x,y,z}
\end{equation}
\begin{equation}
TME(t)=\frac{1}{2}[B'^2_x+B'^2_y+B'^2_z]_{x,y,z}
\end{equation}\\

Taking the inner product of the fluctuation momentum equation \eqref{uprimeequation} with $\bv{u'}$ and of the fluctuation induction equation \eqref{Bprimeequation} with $\bv{B'}$, omitting the stochastic excitation terms, and averaging over the domain, the equations for TKE and TME are:\\

\begin{equation}
\begin{split}
 \frac{d\, TKE}{d t}=&-[\overline{\bv{u'}\cdot(\bv{u'}\cdot\nabla)\overline{\bv{u}}}]_{y,z}+[\overline{\bv{u'}\cdot(\bv{B'}\cdot\nabla)\overline{\bv{B}}}]_{y,z}+[\overline{\bv{u'}\cdot(\overline{\bv{B}}\cdot\nabla)\bv{B'}}]_{y,z}\\
 &- \frac{1}{Re}[\overline{\nabla \bv{u'} : \nabla \bv{u'}}]_{y,z}
\end{split}
\label{TKEeq}
\end{equation}
\begin{equation}
\begin{split}
 \frac{d\, TME}{d t}=&\,[\overline{\bv{B'}\cdot(\bv{B'}\cdot\nabla)\overline{\bv{u}}}]_{y,z}-[\overline{\bv{B'}\cdot(\bv{u'}\cdot\nabla)\overline{\bv{B}}}]_{y,z}+[\overline{\bv{B'}\cdot(\overline{\bv{B}}\cdot\nabla)\bv{u'}}]_{y,z}\\
 &- \frac{1}{Re_m}[\overline{\nabla \bv{B'} : \nabla \bv{B'}}]_{y,z}.
\end{split}
\label{TMEeq}
\end{equation}\\
In \eqref{TKEeq} the first term on the right hand side is the production of TKE by the shear of the mean velocity and the second is the production of TKE by the gradient of the mean magnetic field. In \eqref{TMEeq} the first term is the production of TME by the shear of the mean velocity, which stretches the fluctuation magnetic field, and the second is the production of TME by the gradient of the mean magnetic field. In both equations the fourth term represents dissipation.
The third term in each equation is the field line bending term: the work done by the magnetic tension associated with bending of the mean magnetic field lines, $\overline{\bv{B}}$, by the fluctuations, which mediates Alfv\'enic exchange between fluctuation kinetic and magnetic energy. The bending term in the TME equation is equal and opposite to the bending term in the TKE equation, representing exchange of energy between TKE and TME. The production terms in \eqref{TKEeq} and \eqref{TMEeq} are dominated by production by the background velocity, $[\overline{u}_x]_z$, the streak velocity, $\overline{u}_{xs}$, and the streak magnetic field, $\overline{B}_{xs}$, so that the TKE and TME balances can be written approximately in terms of these production terms, the bending term, $\mathcal{T}$, the bending term arising from the fluctuation-fluctuation nonlinearity, $X$, and dissipation:
\begin{equation}
\frac{d\, TKE}{d t}\approx \mathcal{P}_{[\overline{u}_x]_z\rightarrow TKE}+\mathcal{P}_{\overline{u}_{xs}\rightarrow TKE}+ \mathcal{P}_{\overline{B}_{xs}\rightarrow TKE}+\mathcal{T}+X+ \mathcal{E}_{TKE},
\label{TKEapprox}
\end{equation}
\begin{equation}
\frac{d\, TME}{d t}\approx \mathcal{P}_{[\overline{u}_x]_z\rightarrow TME}+\mathcal{P}_{\overline{u}_{xs}\rightarrow TME}+ \mathcal{P}_{\overline{B}_{xs}\rightarrow TME}-\mathcal{T}-X+ \mathcal{E}_{TME},
\label{TMEapprox}
\end{equation} in which these separate mechanisms have been identified:\\
\begin{equation}
\mathcal{P}_{[\overline{u}_x]_z\rightarrow TKE}=-[\frac{d [\overline{u}_x]_z}{d y}\overline{u'_yu'_x}]_{y,z},
\end{equation}
\begin{equation}
\mathcal{P}_{\overline{u}_{xs}\rightarrow TKE}=-[\frac{\partial \overline{u}_{xs}}{\partial y}\overline{u'_yu'_x}]_{y,z}-[\frac{\partial \overline{u}_{xs}}{\partial z}\overline{u'_zu'_x}]_{y,z},
\end{equation}
\begin{equation}
\mathcal{P}_{\overline{B}_{xs}\rightarrow TKE}=[\frac{\partial \overline{B}_{xs}}{\partial y}\overline{B_y'u'_x}]_{y,z}+[\frac{\partial \overline{B}_{xs}}{\partial z}\overline{B_z'u'_x}]_{y,z}
\end{equation}
\begin{equation}
\mathcal{E}_{TKE}=- \frac{1}{Re}[\overline{\nabla \bv{u'}: \nabla \bv{u'}}]_{y,z} 
\end{equation}
\begin{equation}
\mathcal{P}_{[\overline{u}_x]_z\rightarrow TME}=[\frac{d [\overline{u}_x]_z}{d y}\overline{B_y'B_x'}]_{y,z},
\end{equation}
\begin{equation}
\mathcal{P}_{\overline{u}_{xs}\rightarrow TME}=[\frac{\partial \overline{u}_{xs}}{\partial y}\overline{B_y'B_x'}]_{y,z}+[\frac{\partial \overline{u}_{xs}}{\partial z}\overline{B_z'B_x'}]_{y,z},
\end{equation}
\begin{equation}
\mathcal{P}_{\overline{B}_{xs}\rightarrow TME}=-[\frac{\partial \overline{B}_{xs}}{\partial y}\overline{u_y'B_x'}]_{y,z}-[\frac{\partial \overline{B}_{xs}}{\partial z}\overline{u_z'B_x'}]_{y,z}
\end{equation}
\begin{equation}
\mathcal{E}_{TME}=- \frac{1}{Re_m}[\overline{\nabla \bv{B'}: \nabla \bv{B'}}]_{y,z}, 
\end{equation}
\begin{equation}
\mathcal{T}=[\overline{\bv{u'}\cdot(\overline{\bv{B}}\cdot\nabla)\bv{B'}}]_{y,z}=-[\overline{\bv{B'}\cdot(\overline{\bv{B}}\cdot\nabla)\bv{u'}}]_{y,z},
\label{bendingdef}
\end{equation}
\begin{equation}
X=[\overline{\bv{u'}\cdot(\bv{B'}\cdot\nabla)\bv{B'}}]_{y,z}=-[\overline{\bv{B'}\cdot(\bv{B'}\cdot\nabla)\bv{u'}}]_{y,z}.
\label{Xdef}
\end{equation}\\

These terms are the TKE production from the background velocity, $\mathcal{P}_{[\overline{u}_x]_z\rightarrow TKE}$, and from the streak velocity, $\mathcal{P}_{\overline{u}_{xs}\rightarrow TKE}$, TKE production from the streak magnetic field, $
\mathcal{P}_{\overline{B}_{xs}\rightarrow TKE}$, dissipation of TKE due to viscosity, $\mathcal{E}_{TKE}$, 
TME production from the background velocity, $\mathcal{P}_{[\overline{u}_x]_z\rightarrow TME}$, and from the streak velocity, $\mathcal{P}_{\overline{u}_{xs}\rightarrow TME}$, TME production from the streak magnetic field, $
\mathcal{P}_{\overline{B}_{xs}\rightarrow TME}$, dissipation of TME due to magnetic diffusivity, $\mathcal{E}_{TME}$, the bending term, $\mathcal{T}$, by which magnetic tension associated with the mean magnetic field exchanges energy between TKE and TME, and the corresponding exchange by the fluctuation-fluctuation nonlinearity, $X$. Both $\mathcal{T}$ and $X$ create no fluctuation energy but move it between TKE and TME, and therefore appear with opposite signs in \eqref{TKEapprox} and \eqref{TMEapprox}. The fluctuation-fluctuation exchange, $X$, is retained in the DNS but is absent in the SSD, in which the fluctuation-fluctuation nonlinearity is parameterized by the stochastic excitation.
\begin{figure}
\centering
\begin{subfigure}{0.8\textwidth} \caption{}
\includegraphics[width=\linewidth]{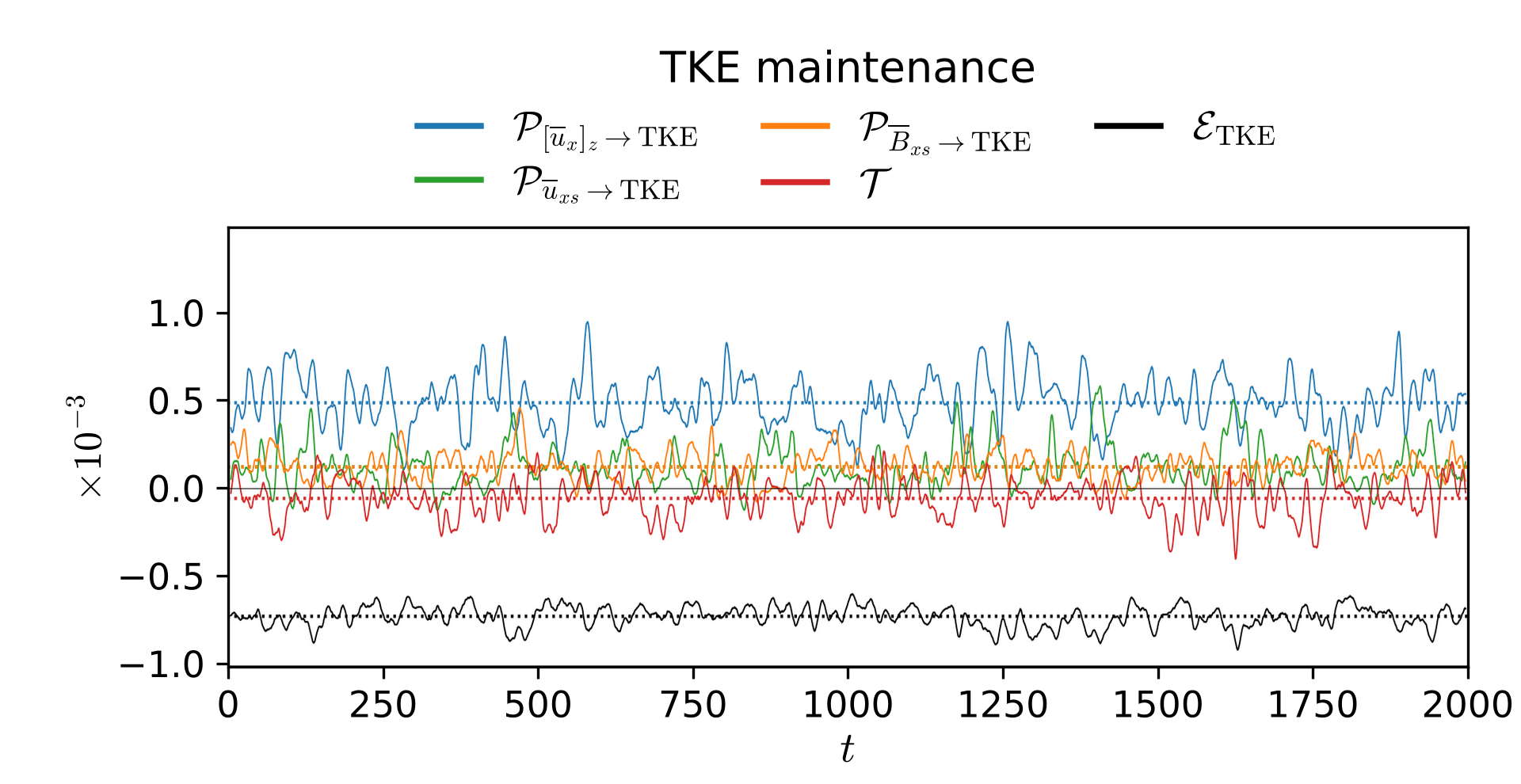}
\end{subfigure}\par
\begin{subfigure}{0.8\textwidth} \caption{}
\includegraphics[width=\linewidth]{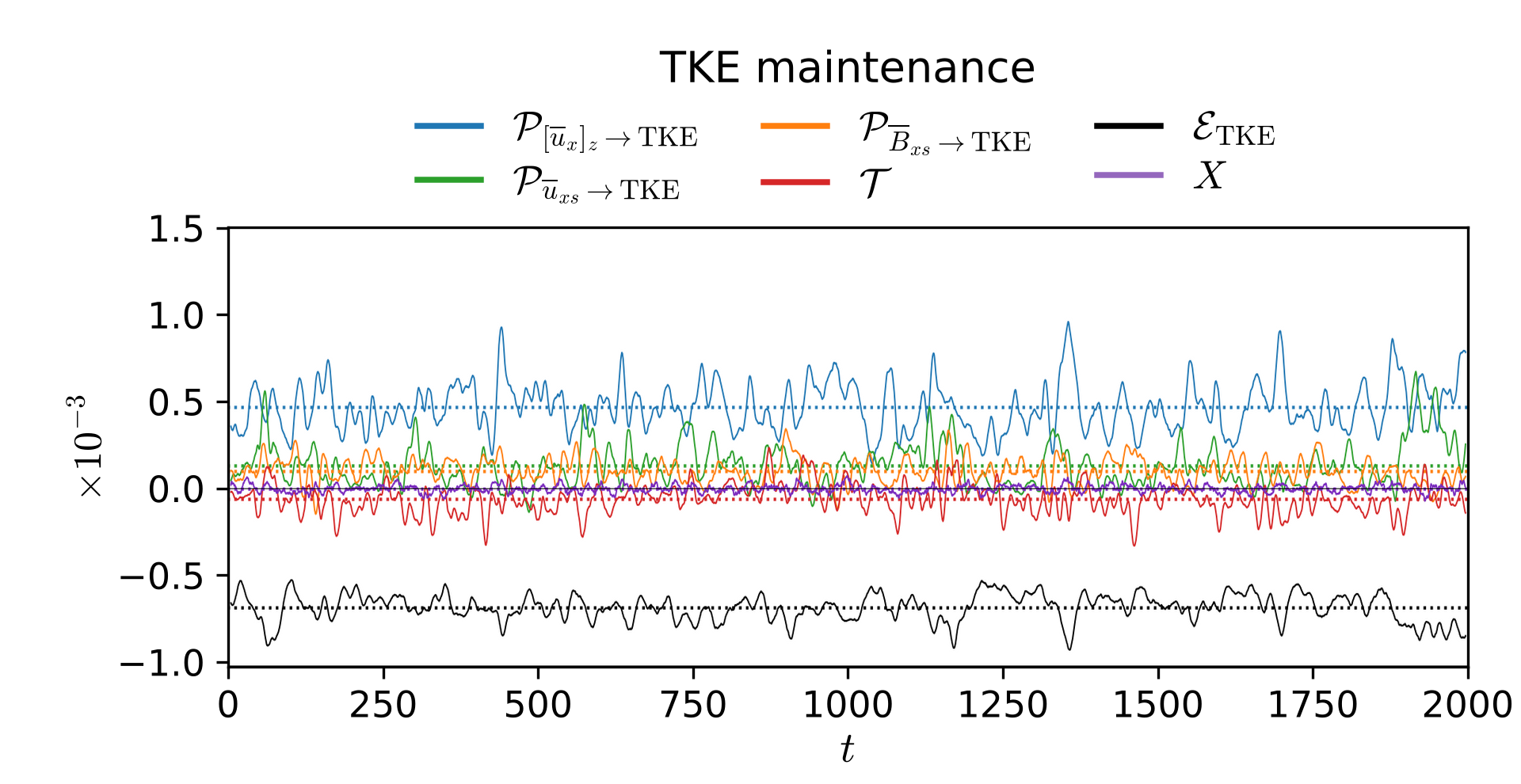}
\end{subfigure}\par
\caption{Maintenance of TKE: (a) SSD with $\epsilon=3.65$ and (b) DNS, both at $Re=Re_m=400$. Shown are TKE production from the background velocity, $\mathcal{P}_{[\overline{u}_x]_z\rightarrow TKE}$, and from the streak velocity, $\mathcal{P}_{\overline{u}_{xs}\rightarrow TKE}$, TKE production from the streak magnetic field, $\mathcal{P}_{\overline{B}_{xs}\rightarrow TKE}$, the bending term, $\mathcal{T}$, the fluctuation-fluctuation exchange, $X$ (DNS only), and dissipation, $\mathcal{E}_{TKE}$.}
\label{TKEplot}
\end{figure}

\begin{figure}
\centering
\begin{subfigure}{0.8\textwidth} \caption{}
\includegraphics[width=\linewidth]{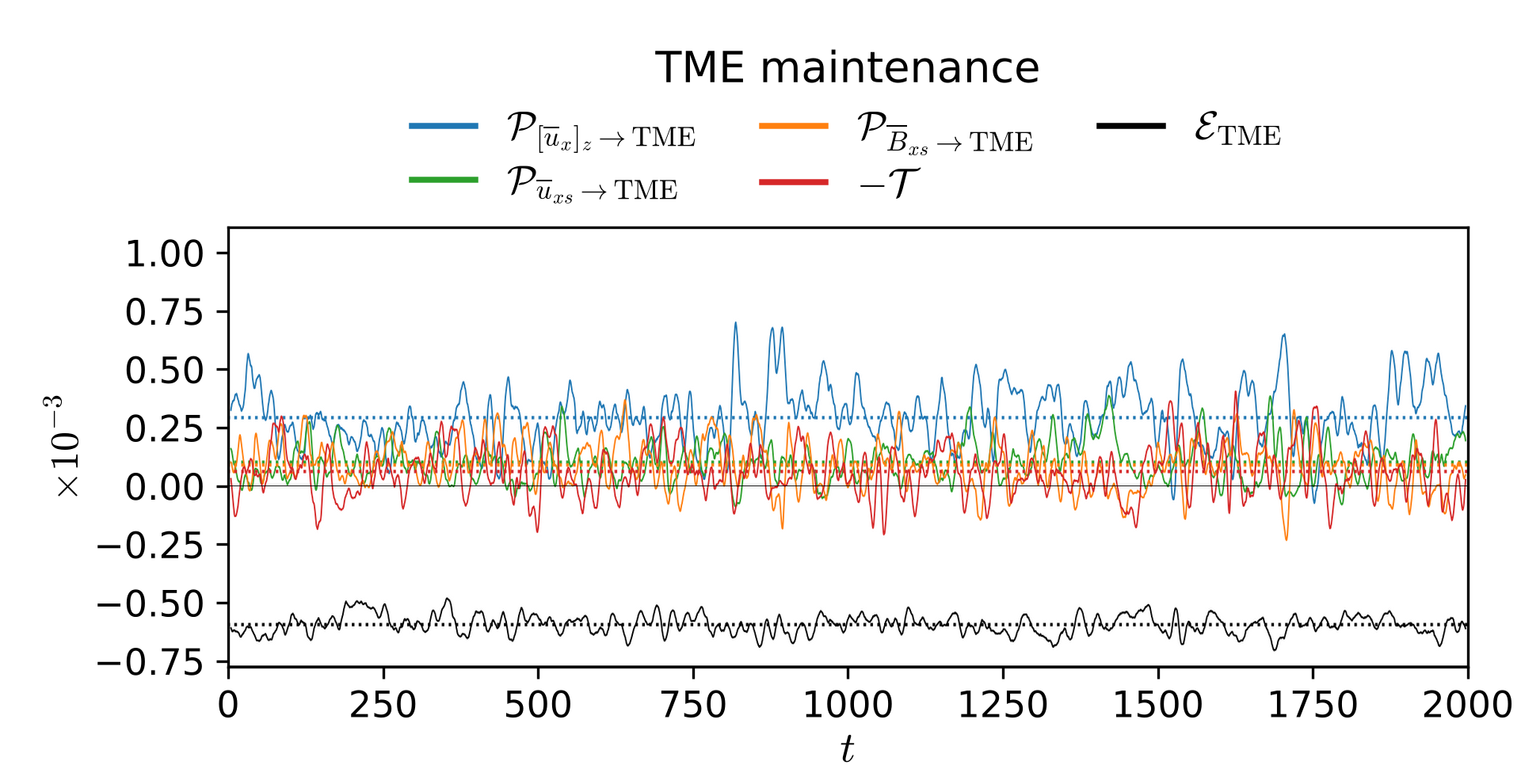}
\end{subfigure}\par
\begin{subfigure}{0.8\textwidth} \caption{}
\includegraphics[width=\linewidth]{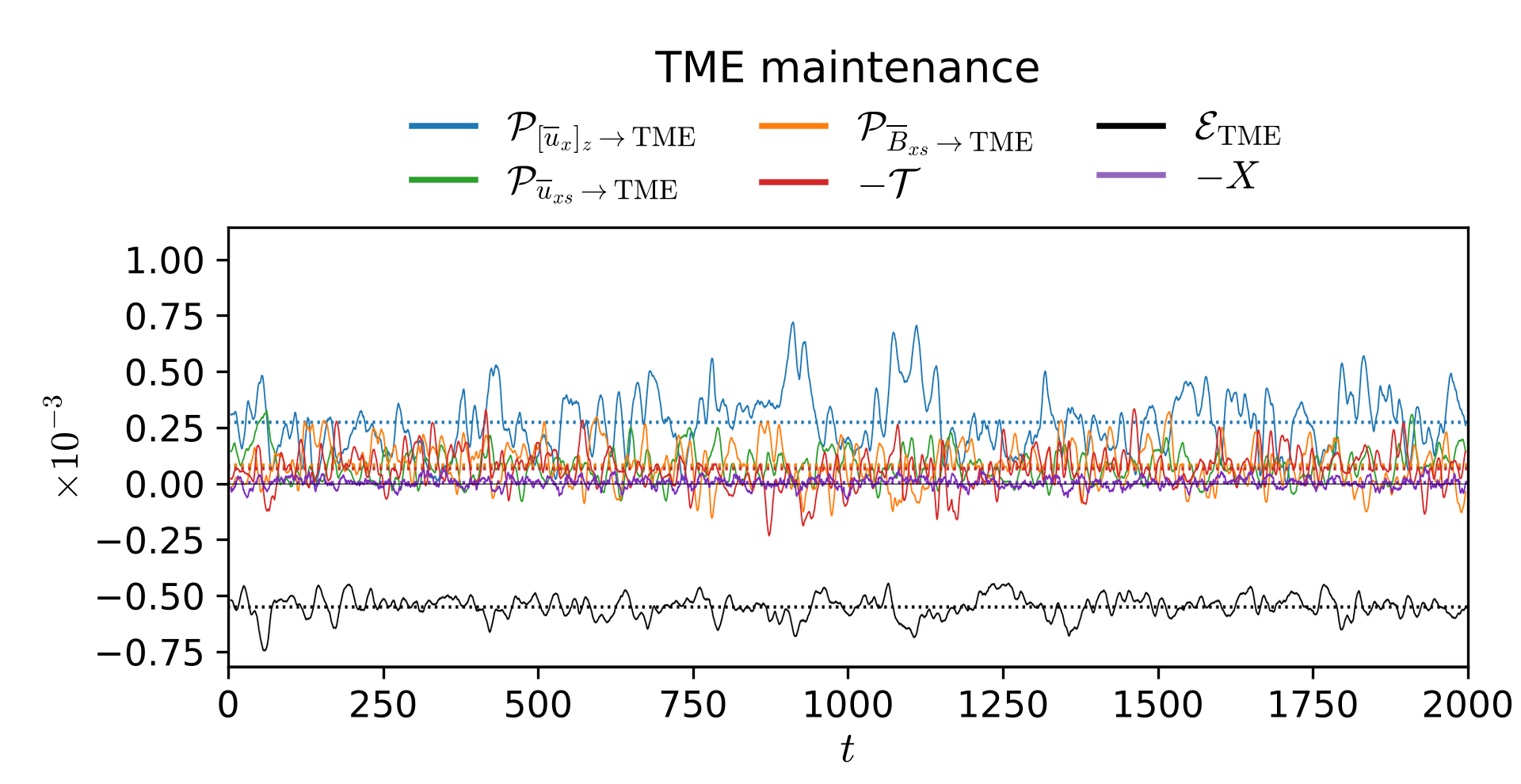}
\end{subfigure}\par
\caption{Maintenance of TME: (a) SSD with $\epsilon=3.65$ and (b) DNS, both at $Re=Re_m=400$. Shown are TME production from the background velocity, $\mathcal{P}_{[\overline{u}_x]_z\rightarrow TME}$, and from the streak velocity, $\mathcal{P}_{\overline{u}_{xs}\rightarrow TME}$, TME production from the streak magnetic field, $\mathcal{P}_{\overline{B}_{xs}\rightarrow TME}$, the bending term, $-\mathcal{T}$, the fluctuation-fluctuation exchange, $-X$ (DNS only), and dissipation, $\mathcal{E}_{TME}$.}
\label{TMEplot}
\end{figure}

Shown in figures \ref{TKEplot} and \ref{TMEplot} are time series of the terms in the TKE and TME balances, respectively, for our case of turbulence at $Re=400$, $Re_m=400$, obtained from the SSD with $\epsilon=3.65$ (panels $a$) and from the DNS (panels $b$). The primary balance for both TKE and TME is between production from the background velocity, $\mathcal{P}_{[\overline{u}_x]_z\rightarrow TKE}$ and $\mathcal{P}_{[\overline{u}_x]_z\rightarrow TME}$, and dissipation, $\mathcal{E}_{TKE}$ and $\mathcal{E}_{TME}$. Significant positive contributions occur from streak velocity, $\mathcal{P}_{\overline{u}_{xs}\rightarrow TKE}$ and $\mathcal{P}_{\overline{u}_{xs}\rightarrow TME}$, and streak magnetic field, $\mathcal{P}_{\overline{B}_{xs}\rightarrow TKE}$ and $\mathcal{P}_{\overline{B}_{xs}\rightarrow TME}$. Results from the SSD and the DNS agree, showing that the TKE and TME are maintained by the same mechanism in both. In particular, the fluctuation-fluctuation exchange, $X$, which is retained in the DNS but absent in the SSD, is negligible (panels $b$ of figures \ref{TKEplot} and \ref{TMEplot}). 

Taken together, the maintenance diagnostics of the RSS components of the mean velocity and magnetic fields (sections \ref{velmaintsec} and \ref{magmaintsec}) and of the TKE and TME (this section) provide a complete account of how this MHD Couette turbulence is maintained. The background shear, $[\overline{u}_x]_z$, maintained by the boundaries, is the ultimate source of energy. The roll velocity, $(\overline{u}_y,\overline{u}_z)$, maintains the streak velocity, $\overline{u}_{xs}$, by lift-up, and the roll magnetic field, $(\overline{B}_y,\overline{B}_z)$, is tilted by the shear to maintain the streak magnetic field, $\overline{B}_{xs}$, by the $\Omega$ mechanism. The streaks in turn shear the background turbulence, systematically organizing the correlations between the velocity and magnetic field fluctuations. The resulting Reynolds and Maxwell stresses drive the roll vorticity, and the resulting EMF drives the roll current, maintaining the roll velocity and roll magnetic field against dissipation and thereby closing the cycle. The fluctuations themselves are maintained by production from the background velocity, the streak velocity and the streak magnetic field, balanced by dissipation, with the bending term exchanging energy between TKE and TME. Each of these components of the mechanism maintaining the turbulence is found to operate in the same manner in the SSD and in the DNS.

\section{Comparison of energy diagnostics between SSD and DNS MHD turbulence}\label{energysec}
\begin{figure}
\centering
\begin{subfigure}{0.8\textwidth} \caption{}
\includegraphics[width=\linewidth]{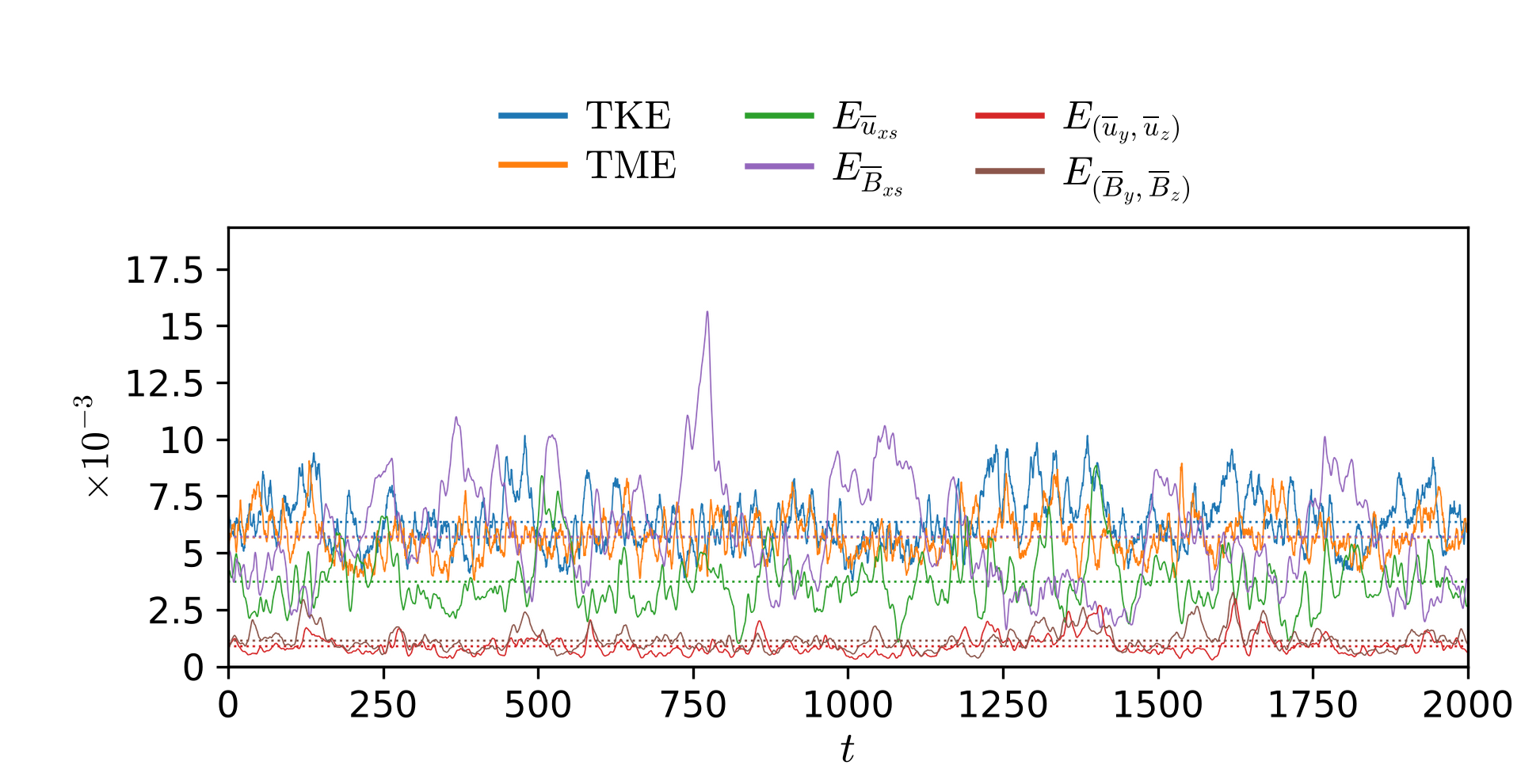}
\end{subfigure}\par
\begin{subfigure}{0.8\textwidth} \caption{}
\includegraphics[width=\linewidth]{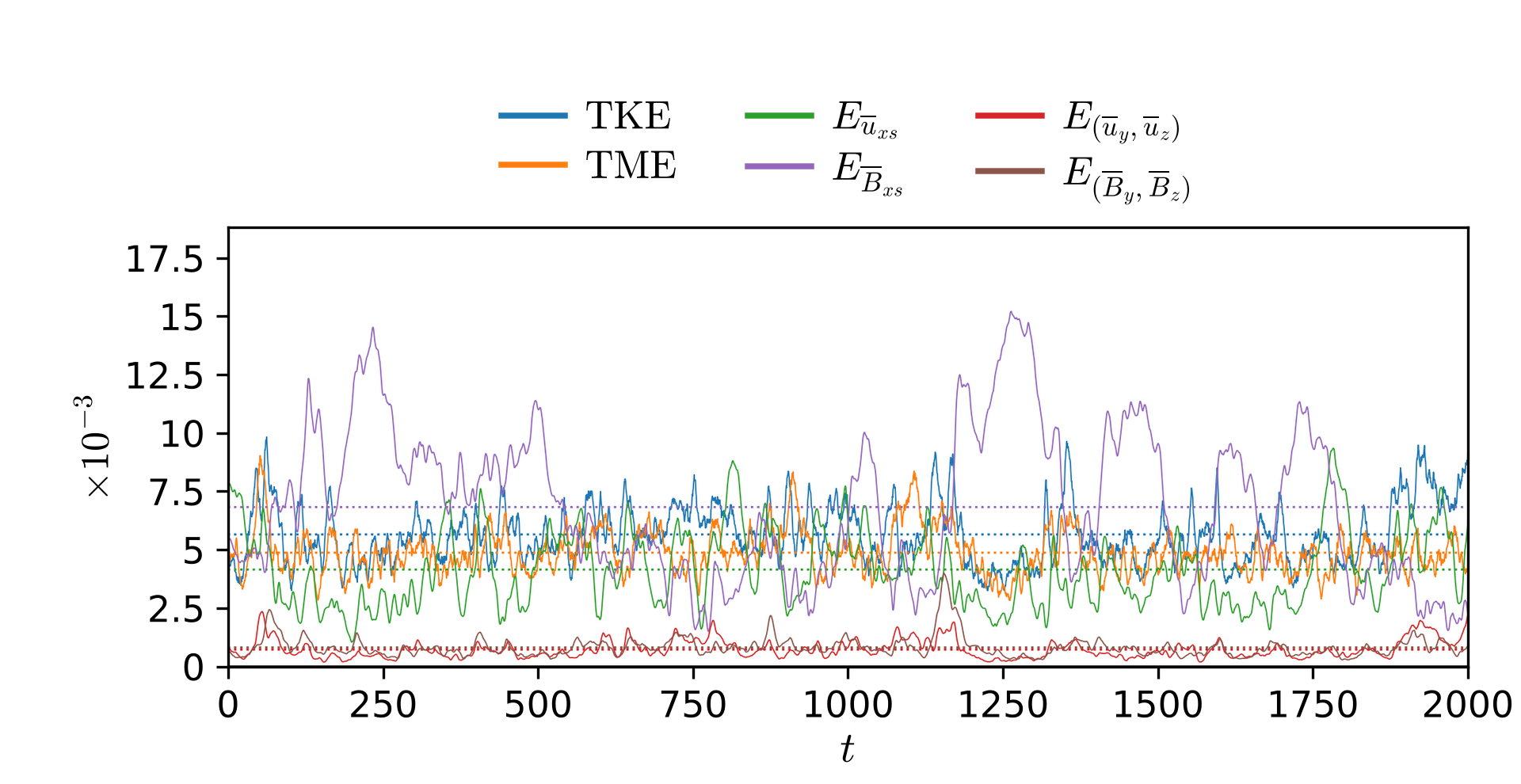}
\end{subfigure}\par
\caption{Time series of the fluctuation kinetic energy, TKE, the fluctuation magnetic energy, TME, the streak and roll kinetic energies, $E_{\overline{u}_{xs}}$ and $E_{(\overline{u}_y,\overline{u}_z)}$, and the streak and roll magnetic energies, $E_{\overline{B}_{xs}}$ and $E_{(\overline{B}_y,\overline{B}_z)}$: (a) SSD with $\epsilon=3.65$ and (b) DNS, both at $Re=Re_m=400$. Dotted lines indicate time means.}
\label{energyplot}
\end{figure}

The structure and maintenance diagnostics presented in the preceding sections show that the turbulence in the SSD and DNS is both similar and maintained by the same mechanisms. To compare the temporal evolution of these turbulent states in more detail, we now consider how energy is partitioned among the mean flow and the fluctuation components of the RSS as a function of time. To this end, in addition to the fluctuation energies, TKE and TME, and the streak magnetic energy, $E_{\overline{B}_{xs}}$, defined in \eqref{EBxsdef}, we define the streak kinetic energy, $E_{\overline{u}_{xs}}$, the roll kinetic energy, $E_{(\overline{u}_y,\overline{u}_z)}$, and the roll magnetic energy, $E_{(\overline{B}_y,\overline{B}_z)}$:
\begin{equation}
E_{\overline{u}_{xs}}=\frac{1}{2}[\overline{u}_{xs}^2]_{y,z},\qquad E_{(\overline{u}_y,\overline{u}_z)}=\frac{1}{2}[\overline{u}_y^2+\overline{u}_z^2]_{y,z},\qquad E_{(\overline{B}_y,\overline{B}_z)}=\frac{1}{2}[\overline{B}_y^2+\overline{B}_z^2]_{y,z}.
\label{Kdef}
\end{equation}
Shown in figure \ref{energyplot} are time series of the fluctuation energies, TKE and TME, together with the energies of the streak and roll components of the mean fields, $E_{\overline{u}_{xs}}$, $E_{(\overline{u}_y,\overline{u}_z)}$, $E_{\overline{B}_{xs}}$ and $E_{(\overline{B}_y,\overline{B}_z)}$, defined in \eqref{EBxsdef} and \eqref{Kdef}, obtained from the SSD with $\epsilon=3.65$ (panel $a$) and from the DNS (panel $b$). In both the SSD and the DNS the streak energies, $E_{\overline{u}_{xs}}$ and $E_{\overline{B}_{xs}}$, substantially exceed the roll energies, $E_{(\overline{u}_y,\overline{u}_z)}$ and $E_{(\overline{B}_y,\overline{B}_z)}$, and the streak magnetic energy, $E_{\overline{B}_{xs}}$, is the largest of the mean-field energies, comparable to the fluctuation energies TKE and TME. Each of these quantities is highly time dependent. Compared with the DNS, the SSD supports larger time-mean roll energies, both kinetic, $E_{(\overline{u}_y,\overline{u}_z)}$, and magnetic, $E_{(\overline{B}_y,\overline{B}_z)}$, but smaller streak energies, both kinetic, $E_{\overline{u}_{xs}}$, and magnetic, $E_{\overline{B}_{xs}}$, together with larger fluctuation energies, TKE and TME. The larger fluctuation energies in the SSD are consistent with the absence of the fluctuation-fluctuation nonlinearity, and therefore of the turbulent cascade, in the SSD: in the DNS the cascade transfers fluctuation energy to small scales, where it is removed by dissipation, whereas in the SSD this route is absent and the fluctuations equilibrate at larger amplitude. The larger fluctuation amplitude in the SSD also explains the larger roll energies: the Reynolds and Maxwell stresses and the EMF arising from the systematically organized fluctuations, which drive the roll vorticity and roll current, are correspondingly stronger, so that the rolls are driven more strongly. This is also reflected in the maintenance diagnostics of the roll vorticity and the roll current density (figures \ref{Omegarturbulence} and \ref{OmegaBrmaintenance}), in which the driving by the fluctuations, the Reynolds stress and Maxwell stress terms, $I_G$ and $I_I$, in the roll vorticity balance and the fluctuation-fluctuation tilting and advection terms, $I_{BG}$ and $I_{BI}$, in the roll current balance, is larger in the SSD than in the DNS. The stronger rolls in the SSD in turn produce greater lift-up of the streak velocity, $I_A$, and greater mean tilting of the roll magnetic field into the streak magnetic field, $I_{BA}$, as seen in the larger amplitudes of these terms in the streak velocity and streak magnetic field maintenance diagnostics for the SSD compared with the DNS (figures \ref{Usturbulence} and \ref{Bsmaintenance}). However, the larger amplitude fluctuations in the SSD also strain the streaks more strongly, as reflected in the larger amplitude negative fluctuation terms in the maintenance of both the streak velocity, $I_B$ and $I_D$, and the streak magnetic field, $I_{BB}$ and $I_{BD}$. Combined together, these effects produce a streak velocity and streak magnetic field with smaller energies, $E_{\overline{u}_{xs}}$ and $E_{\overline{B}_{xs}}$, in the SSD than in the DNS, despite the greater lift-up and mean tilting driven by the stronger rolls. Apart from these differences in the means of their energy diagnostics, which are analogous to those seen in a study of Poiseuille flow turbulence \citep{Farrell 2016}, turbulence in the SSD and in the DNS exhibit similar temporal variability.

\FloatBarrier
\section{Discussion and Conclusion}
The MHD equations of Couette flow formulated as a perturbation stability problem about a laminar velocity state do not support  instability, much less an unstable RSS with large scale coherent velocity and magnetic fields. Although  the RSS is ubiquitous in both turbulent and transitional shear flows and DNS experiments confirmed that it is required for maintaining turbulence in shear flow, the existence of an instability with RSS form and a dynamical system in which this structure maintains a statistically stable  turbulent state in the NS eluded explicit dynamical analysis in both conducting and nonconducting shear flow. The reason why the dynamics of the RSS remained obscure is that this structure and the instability forming it are nonlinear.  The key to discovering this foundational instability, and the pivotal dynamics of the RSS in general, lies in analyzing the turbulence dynamics using an SSD formulation of the NS. In the S3T SSD closed at second order the second cumulant, which is the ensemble mean covariance of the state fluctuations, is treated as a linear variable in the perturbation analysis, although it is quadratically nonlinear as a state variable, which allows the full power of linear perturbation analysis to be brought to bear on this nonlinear instability.

Previously,  SSD had been used to study the formation of the RSS as an unstable mode and its maintenance as a stable fixed point as well as the equilibration of the RSS to self-sustaining turbulent states in nonconducting wall bounded shear flow \citep{Farrell-Ioannou-2016-bifur, Farrell-Ioannou-2012}. 
 The similarity of the NS and induction equations motivated the extension of these previous results to the study of MHD turbulence in conducting  Couette flow using SSD. Further motivation came from the fact that the RSS in shear flow supports a poloidal velocity field with a helical structure that suggests a natural support for maintaining an associated toroidal magnetic field.  In this work we found that, in the presence of a background field of small amplitude random turbulence in a conducting Couette flow, RSS emerge as unstable modes with remarkably similar structure in the velocity field and the magnetic field---almost as if these two RSS coexisted with a $90^\circ$ phase shift separating them in $z$.  Moreover, we found that these RSS structures are synergistic in the sense of cooperating in the formation and growth of the instability although the ultimate energy source for the RSS is the velocity shear at the boundary. 
 Because the SSD we use is the NS equations closed at second order, it contains the nonlinear mechanism responsible for establishing finite amplitude equilibrium states, both for fixed points and, in the case of turbulent RSS states, the statistical mean state. We identified equilibrium regimes consisting of stable ($\bv{\overline{u}}$ and $\bv{\overline{B}})$ fixed points containing both fields at finite amplitude, excited turbulent states, self sustaining turbulent states supporting  $\bv{\overline{u}}$ only, and self sustaining  turbulent states containing both ($\bv{\overline{u}}$ and $\bv{\overline{B}})$. 
These self sustaining states emerge after transition to RSS turbulence has occurred.  They are revealed by eliminating the stochastic turbulence excitation from the SSD.  If, when this is done, the turbulence transitions to a statistically stable turbulent state, we conclude that a self sustaining state is supported for the parameters used. 

In the self sustaining states,
maintenance of a coherent large scale RSS supporting only a mean $\overline{\bv{u}}$ field occurs for some parameters.  For other parameters both the $\overline{\bv{u}}$ and  $\overline{\bv{B}}$ field are supported in the RSS, which corresponds to a dynamo. Using the magnetic Prandtl number, $Pr_{m}$, as a parameter we found that transition from a self sustaining state in the velocity field only to a self
sustaining state in both the velocity and the magnetic field, corresponding to a dynamo, manifested as an abrupt  bifurcation in the statistical state occurring at a critical  $Pr_{mc}$.

Using the S3T SSD we are able to obtain and analyze the unstable RSS modes and the nonlinear equilibria proceeding from these modes, both fixed point and turbulent.  This allowed us to study in detail the physical mechanisms underlying the RSS dynamics.

We found that the mechanism supporting the streak and the roll component of the velocity field of the turbulent MHD RSS is similar to that previously found to support the streak and the roll velocity component of the  RSS in the SSD for non-conducting Couette flow. Dominant balance for the streak velocity in both cases is between positive contributions from lift-up and negative contributions from dissipation and fluctuation-fluctuation terms. In turbulent MHD RSS, the difference is that small Maxwell stress components contribute additional negative forcing to maintaining $\overline{u}_{xs}$.

Similar to the roll vorticity force balance in nonconducting wall bounded shear
flow, the dominant balance supporting the roll velocity component of the MHD RSS is between positive contributions from fluctuation-fluctuation terms  opposed
by a negative contribution from dissipation. In turbulent MHD RSS, the
difference is that the positive contributions from the fluctuation-fluctuation terms come from
not only the Reynolds stress term, but also the Maxwell stress term which is indicative of a synergy between the velocity and magnetic field components in maintaining the combined RSS.

In the case of the magnetic field streak component of the RSS, $\overline{B}_{xs}$, we found that 
 the dominant balance is between a positive contribution from mean tilting (which is analogous to lift-up in the maintenance of the streak velocity component), opposed by negative contributions from dissipation and fluctuation-fluctuation terms. The correlation between the fluctuation velocity field and the fluctuation magnetic field results in systematic extraction of energy from the streak  magnetic field, which energy is transferred to the fluctuations.

Turning to the maintenance of the roll magnetic field component, which is the current density in the $x$ direction, we find that  fluctuation tilting and stretching, and fluctuation advection, contribute positively to maintaining the current density of the roll while dissipation, contributes negatively.  In summary,
in both the maintenance of the streak velocity and the streak magnetic field, fluctuation-fluctuation terms contribute negatively. Similarly, in both the maintenance of the roll velocity and the roll magnetic field, it is the fluctuation-fluctuation terms that contribute positively.  This result demonstrates the centrality of fluxes arising from the correlation of turbulent fluctuations in the velocity and magnetic field in sustaining both the roll velocity and the roll (or poloidal) magnetic field that constitutes the fluctuation feedback allowing the $\Omega$ dynamo to be destabilized and maintained. This feedback arises from the electromotive force produced by the systematic correlation of the velocity and magnetic field fluctuations organized by the RSS. Although this feedback has an effect analogous to that of the $\alpha$ effect, in that it circumvents Cowling's theorem by regenerating the poloidal field, it is mechanistically distinct from the $\alpha$ effect.

\medskip
\noindent\textbf{Declaration of interests.} {\small The authors report no conflict of interest.\par}

\appendix
\section{Sign-weighted maintenance diagnostics}\label{app:Iterms}
This appendix defines the sign-weighted diagnostics of the physically distinct terms in the balance equations for the streak velocity, the roll vorticity, the streak magnetic field and the roll current density discussed in \S\S\,8--9. Because streaks and rolls of both signs occur, each term is multiplied by the sign of the local streak or roll component, so that a positive value indicates that the term reinforces that component where it acts.

\subsection{Streak velocity}
The terms in the dynamics maintaining $\overline{u}_{xs}$ are:
\begin{equation}I_A=sign(\overline{u}_{xs})\cdot(-(\overline{u}_y(\partial_y \overline{u}_x)-[\overline{u}_y(\partial_y \overline{u}_x)]_z)-(\overline{u}_z(\partial_z \overline{u}_x)-[\overline{u}_z(\partial_z \overline{u}_x)]_z))\end{equation}
\begin{equation}
I_B=sign(\overline{u}_{xs})\cdot(-([u_y'(\partial_y u_x')]_{x}-[u_y'(\partial_y u_x')]_{x,z})-([u_z'(\partial_z u_x')]_{x}-[u_z'(\partial_z u_x')]_{x,z}))
\end{equation}
\begin{equation}
I_C=sign(\overline{u}_{xs})\cdot ( (\overline{B}_y(\partial_y\overline{B}_x)-[\overline{B}_y(\partial_y\overline{B}_x)]_{z})+(\overline{B}_z(\partial_z\overline{B}_x)-[\overline{B}_z(\partial_z\overline{B}_x)]_{z}))
\end{equation}

\begin{equation}
I_D= sign(\overline{u}_{xs})\cdot (([B'_y(\partial_yB'_x)]_{x}-[B'_y(\partial_yB'_x)]_{x,z})+([B'_z(\partial_zB'_x)]_{x}-[B'_z(\partial_zB'_x)]_{x,z}))
\end{equation}
\begin{equation}
I_E=sign(\overline{u}_{xs})\cdot(\frac{1}{Re}\Delta_1 \overline{u}_{xs})
\end{equation}\\

\subsection{Roll vorticity}
The terms in the dynamics maintaining $\overline{\omega}_{xr}$ are:
\begin{equation}
I_F=sign(\overline{\omega}_{xr})\cdot (-((\overline{u}_y\partial_y + \overline{u}_z \partial_z)\overline{\omega}_{x}-[(\overline{u}_y\partial_y + \overline{u}_z \partial_z) \overline{\omega}_{x}]_z))
\end{equation}
\begin{equation}
I_G=sign(\overline{\omega}_{xr}) \cdot (\partial_{zz}-\partial_{yy})([u_y'u_z']_x-[u_y'u_z']_{x,z})- \partial_{yz}([u_z'u_z']_x-[u_y'u_y']_x)
\end{equation}

\begin{equation}
I_H=sign(\overline{\omega}_{xr}) \cdot ((\partial_{yy}-\partial_{zz})(\overline{B}_y\overline{B}_z-[\overline{B}_y\overline{B}_z]_z)+\partial_{yz}(\overline{B}_z\overline{B}_z-\overline{B}_y\overline{B}_y))
\end{equation}
\begin{equation}
I_I=sign(\overline{\omega}_{xr})\cdot ((\partial_{yy}-\partial_{zz})([B'_yB'_z]_x-[B'_yB'_z]_{x,z})+\partial_{yz}([B'_zB'_z]_x-[B'_yB'_y]_x))
\end{equation}
\begin{equation}
I_J=sign(\overline{\omega}_{xr}) \cdot \Delta_1\frac{\overline{\omega}_{xr}}{Re}
\end{equation}\\

\subsection{Streak magnetic field}
The terms in the dynamics maintaining $\overline{B}_{xs}$ are:
\begin{equation}
I_{BA}=sign(\overline{B}_{xs})\cdot (((\partial_y \overline{u}_x)\overline{B}_y-[(\partial_y \overline{u}_x)\overline{B}_y]_{z})+((\partial_z\overline{u}_x)\overline{B}_z-[(\partial_z\overline{u}_x)\overline{B}_z]_z))
\end{equation}
\begin{equation}
\begin{split}
I_{BB}=sign(\overline{B}_{xs})\cdot (([(\partial_x u'_x)B'_x]_x-[(\partial_x u'_x)B'_x]_{x,z})+([(\partial_y u'_x)B'_y]_x-[(\partial_y u'_x)B'_y]_{x,z})\\+([(\partial_z u'_x)B'_z]_x-[(\partial_z u'_x)B'_z]_{x,z}))
\end{split}
\end{equation}
\begin{equation}
I_{BC}=sign(\overline{B}_{xs})\cdot (-((\partial_y \overline{B}_x)\overline{u}_y-[(\partial_y \overline{B}_x)\overline{u}_y]_{z})\\-((\partial_z \overline{B}_x)\overline{u}_z-[(\partial_z \overline{B}_x)\overline{u}_z]_z))
\end{equation}
\begin{equation}
\begin{split}
I_{BD}=sign(\overline{B}_{xs})\cdot (-([(\partial_x B'_x)u'_x]_x-[(\partial_x B'_x)u'_x]_{x,z})-([(\partial_y B'_x)u'_y]_x-[{(\partial_y B'_x)u'_y}]_{x,z})\\-([(\partial_z B'_x)u'_z]_x-[{(\partial_z B'_x)u'_z}]_{x,z}))
\end{split}
\end{equation}
\begin{equation}
I_{BE}=sign(\overline{B}_{xs})\cdot \frac{1}{Re_{m}}\Delta_1 \overline{B}_{xs}
\end{equation}

\subsection{Roll current density}
The terms in the dynamics maintaining $\overline{\zeta}_{xr}$ are:
\begin{equation}
\begin{split}
I_{BF}=sign(\overline{\zeta}_{xr})\cdot (\partial_y((\partial_y\overline{u}_z)\overline{B}_y-[(\partial_y\overline{u}_z)\overline{B}_y]_z)+\partial_y((\partial_z\overline{u}_z)\overline{B}_z-[(\partial_z \overline{u}_z) \overline{B}_z]_{z})\\-\partial_{z}((\partial_y \overline{u}_y) \overline{B}_y)-\partial_z((\partial_z \overline{u}_y) \overline{B}_z))
\end{split}
\end{equation}
\begin{equation}
\begin{split}
I_{BG}=sign(\overline{\zeta}_{xr})\cdot (\partial_y([ (\partial_x u'_z)B'_x]_x-[(\partial_x u'_z)B'_x]_{x,z})+\partial_y([ (\partial_y u'_z)B'_y]_x-[ (\partial_y u'_z)B'_y]_{x,z})\\+\partial_y([ (\partial_z u'_z )B'_z]_x-[ (\partial_z u'_z )B'_z]_{x,z})-\partial_{z}([ (\partial_x u'_y) B'_x]_x)-\partial_z([ (\partial_y u'_y) B'_y]_x)\\-\partial_z(  [(\partial_z u'_y) B'_z]_x))
\end{split}
\end{equation}

\begin{equation}
\begin{split}
I_{BH}=sign(\overline{\zeta}_{xr}) \cdot (\partial_z((\partial_y \overline{B}_y)\overline{u}_y)+\partial_z((\partial_z\overline{B}_y)\overline{u}_z)-\partial_y( (\partial_y\overline{B}_z)\overline{u}_y-[(\partial_y\overline{B}_z)\overline{u}_y]_z)\\-\partial_y((\partial_z\overline{B}_z)\overline{u}_z-[(\partial_z\overline{B}_z)\overline{u}_z]_{z}))
\end{split}
\end{equation}

\begin{equation}
\begin{split}
I_{BI}=sign(\overline{\zeta}_{xr})\cdot (\partial_z([ (\partial_x B'_y)u'_x]_{x})+\partial_z([ (\partial_y B'_y)u'_y]_{x})+\partial_z([(\partial_z B'_y)u'_z]_x)\\-\partial_y([(\partial_x B'_z)u'_x]_x-[(\partial_x B'_z)u'_x]_{x,z})-\partial_y([(\partial_y B'_z)u'_y]_x-[(\partial_y B'_z)u'_y]_{x,z})\\-\partial_y([(\partial_z B'_z)u'_z]_x-[(\partial_z B'_z)u'_z]_{x,z}))
\end{split}
\end{equation}
\begin{equation}
I_{BJ}= sign(\overline{\zeta}_{xr}) \cdot (\frac{1}{Re_{m}}\Delta_1 \overline{\zeta}_{xr})
\end{equation}

\end{document}